\documentclass[twocolumn]{autart}

\usepackage{cite}
\usepackage{amsmath,amssymb,amsfonts}
\usepackage{graphicx}
\usepackage{mathtools}
\usepackage{algorithm,algorithmicx}
\usepackage{algpseudocode}
\usepackage{textcomp}

\usepackage{tikz}
\usetikzlibrary{arrows,shapes,trees,calc,positioning}

\DeclareMathAlphabet{\pazocal}{OMS}{zplm}{m}{n}

\newcommand{\norm}[1]{\left\lVert#1\right\rVert}
\newcommand{\col}[1]{\mathrm{col}\left(#1\right)}
\newcommand{\cov}[1]{\mathrm{cov}\left(#1\right)}
\newcommand{\vect}[1]{\mathrm{vec}\left(#1\right)}
\newcommand{\tr}[1]{\mathrm{tr}\left(#1\right)}
\newcommand{\pr}[1]{\mathrm{Pr}\left(#1\right)}
\newcommand{\diag}[1]{\mathrm{diag}\left(#1\right)}
\newcommand{\blkdiag}[1]{\mathrm{blkdiag}\left(#1\right)}
\newcommand{\logdet}[1]{\mathrm{logdet}\left(#1\right)}
\newcommand{\detnew}[1]{\mathrm{det}\left(#1\right)}
\newcommand{\nullsp}[1]{\mathrm{null}\left(#1\right)}

\newtheorem{remark}{Remark}

\newtheorem{assumption}{Assumption}

\newtheorem{corollary}{Corollary}
\newtheorem{lemma}{Lemma}
\usepackage[T1]{fontenc}

\begin{document}

\begin{frontmatter}

\title{Data-Driven Prediction and Control of Hammerstein-Wiener Systems with Implicit Gaussian Processes\thanksref{footnoteinfo}}

\thanks[footnoteinfo]{This paper was not presented at any IFAC 
meeting. This work was supported by the Lower Saxony Ministry for Science and Culture within the program zukunft.niedersachsen. Corresponding author Mingzhou Yin. Tel. +49 511 762 18903. 
Fax +49 511 762 4536.}

\author[irt]{Mingzhou Yin}\ead{yin@irt.uni-hannover.de},
\author[irt]{Matthias A. M\"uller}\ead{mueller@irt.uni-hannover.de}

\address[irt]{Institute of Automatic Control, Leibniz University Hannover, 30167 Hannover, Germany}


\begin{abstract}
This work investigates data-driven prediction and control of Hammerstein-Wiener systems using physics-informed Gaussian process (GP) models that encode the block-oriented model structure. Data-driven prediction algorithms have been developed for structured nonlinear systems based on Willems' fundamental lemma. However, existing frameworks do not apply to output nonlinearities in Wiener systems and rely on a finite-dimensional dictionary of basis functions for Hammerstein systems. In this work, an implicit predictor structure is considered, leveraging the linearity for the dynamical part of the model. This implicit function is learned by GP regression, utilizing carefully designed structured kernel functions from linear model parameters and GP priors for the nonlinearities. Virtual derivative points are added to the regression by expectation propagation to encode monotonicity information of the nonlinearities. The linear model parameters are estimated as hyperparameters by assuming a stable spline hyperprior. The implicit GP model provides explicit output prediction by optimizing selected optimality criteria. The implicit model is also applied to receding horizon control with the expected control cost and chance constraint satisfaction guarantee. Numerical results demonstrate that the proposed prediction and control algorithms are superior to black-box GP models without model structure knowledge.
\end{abstract}

\end{frontmatter}

\section{Introduction}
\label{sec:1}

In recent years, the problem of data-driven prediction has drawn significant attention due to the emergence of more complex systems and the availability of big data. This problem aims to obtain nonparametric predictors directly from collected input-output data using minimal model knowledge of the system. In this way, model-based predictive controllers can be readily adapted into data-driven ones by employing data-driven predictors. This work investigates data-driven prediction and control of Hammerstein-Wiener (H-W) systems. H-W systems are nonlinear block-oriented models with static input and output nonlinearities \cite{giri2010block}, which are widely used to model, e.g., chemical processes \cite{Abinayadhevi_2015}, electromechanical systems \cite{Zhang_2019}, and ionospheric dynamics \cite{Palanthandalam_Madapusi}. More generally, the input and output nonlinearities can be used to model actuator and sensor nonlinearities, respectively.

Data-driven prediction can be seen as finding a mapping from inputs and initial conditions to outputs, where any function approximator can be applied to learn the mapping. In this regard, a basis function decomposition approach is used in \cite{Lazar_2024}. Gaussian process (GP) or kernel regression has been widely used in system identification to estimate the model \cite{pillonetto2022regularized,Pillonetto_2014} by designing kernel structures for stable nonlinear systems inspired by system theory \cite{Pillonetto_2011}. Kernel-based identification for Hammerstein and Wiener systems is investigated separately in \cite{Risuleo_2017,Risuleo_2019} for one-step-ahead prediction. This idea of using system-theoretic knowledge (such as model structure) in learning is sometimes known as physics-informed approaches \cite{Cross_2024}. Predictions from the GP model can then be used in receding horizon control. This approach is referred to as Gaussian process model predictive control (GP-MPC) \cite{1383790,Hewing_2020,Bradford_2020,Maiworm_2021,scampicchio2025gaussian,prajapat2025finite}. However, existing GP-MPC schemes employ either a fully black-box model without any model structure or an additive model structure, consisting of a known nominal model plus black-box dynamics, neither of which can encode the H-W structure to obtain a physics-informed model. In addition, these schemes only learn a one-step-ahead predictor and require uncertainty propagation over the prediction horizon, which is challenging and often requires approximation \cite{Hewing_2020,scampicchio2025gaussian}.

On the other hand, following the seminal result in \cite{Willems_2005}, known as Willems' fundamental lemma (WFL), data-driven prediction of linear systems has been widely studied by finding suitable linear combinations of persistently exciting data, subject to input and initial condition constraints \cite{Markovsky_20080}. The predictor can be reformulated as a multi-step-ahead autoregressive with extra input (ARX) structure. Applying the predictor to receding horizon control leads to data-driven (or data-enabled) predictive control (DDPC), for which multiple successful algorithms have been proposed, e.g., \cite{Coulson_2019_reg,Berberich_2021,Yin_2023,Breschi_2023,alsalti2024robust,Chiuso_2025}.

Extensions of these linear algorithms to nonlinear systems have been analyzed for specific nonlinear model classes, including Hammerstein systems \cite{Berberich_2020,Markovsky_2023,Molodchyk_2024}, second-order Volterra systems \cite{Rueda_Escobedo_2020}, feedback linearizable systems \cite{Alsalti_2023,Molodchyk_2024}, bilinear systems \cite{Yuan_2022,Markovsky_2023}, and systems with finite-dimensional Koopman embedding \cite{shang2024willems}. These extensions generalize WFL by applying generalized signal embedding to the signal matrix. Some also apply a kernelized version of WFL by considering the inner product of the signal embedding using the ``kernel trick'' \cite{Molodchyk_2024,pmlr-v144-lian21a,Huang_2024}. However, for H-W systems, existing works \cite{Berberich_2020,Markovsky_2023,Molodchyk_2024} address only the input nonlinearity (the Hammerstein part) but not the output nonlinearity (the Wiener part)\footnote{Wiener systems are discussed in \cite{Berberich_2020}, but an extension of WFL that enables data-driven prediction is not obtained for this system class, cf. Section~\ref{sec:2b} below.}. In addition, the signal embedding or its kernel reformulation requires a known finite-dimensional dictionary of basis functions to satisfy the persistency of excitation condition. Such information can be impractical to obtain. The lifted signal space also makes it difficult to satisfy the persistency of excitation condition, requiring a long input sequence with tailored design \cite{Alsalti_2023_2}.

In this work, data-driven prediction of H-W systems is first formulated as an implicit function learning problem by considering the multi-step-ahead ARX structure for the linear part. GP priors are assumed for the nonlinearities, 
which correspond to an infinite-dimensional reproducing kernel Hilbert space (RKHS), where universal approximation capability can be attained for common designs (such as squared exponential kernels) without any knowledge about the nonlinearities \cite{JMLR:v7:micchelli06a}. Then, the structured kernel function for the implicit prediction model is derived together with the linear model parameters. Compared with an explicit black-box GP model with general kernel functions, the structured kernel design provides a physics-informed design by considering a smaller function space for the predictor that is compatible with the H-W model structure.

To retain the well-definedness of the output nonlinearities in the implicit GP model, virtual derivative points are added to promote monotonic output nonlinearities by expectation propagation \cite{10.7551/mitpress/3206.001.0001,pmlr-v9-riihimaki10a}. The implicit GP model \cite{Martens_2017} provides explicit posterior predictions by optimizing selected optimality criteria, such as maximum likelihood. The linear model parameters act as hyperparameters in the proposed GP model, which are estimated by enforcing a stable spline hyperprior and solving a joint maximum-a-posteriori/maximum-likelihood (JMAP-ML) problem to avoid overfitting \cite{pillonetto2022regularized,Pillonetto_2014}. The nonlinearities can also be recovered separately using additional GP regression procedures.

The implicit GP predictor is applied to receding horizon control by considering the expected control cost and guaranteeing output chance constraint satisfaction. The proposed predictor provides multi-step-ahead predictions by nature, circumventing the difficult problem of uncertainty propagation in standard GP-MPC with one-step-ahead prediction. Finally, the predictor and the control algorithm are verified in numerical examples and demonstrate better performance than the black-box GP predictor with a general nonlinear ARX model.

The remainder of the paper is organized as follows. Section~\ref{sec:2} introduces the problem and presents the implicit predictor structure. Section~\ref{sec:3} proposes the data-driven prediction algorithm by implicit GP regression. Section~\ref{sec:4} applies the data-driven predictor to receding horizon control, followed by numerical examples in Section~\ref{sec:5}. Section~\ref{sec:6} concludes the paper.

\textit{Notation.} The expected value and covariance of a random variable are denoted by $\mathbb{E}[\cdot]$ and $\cov{\cdot}$, respectively. The symbol $\pr{\cdot}$ indicates the probability of a random event. A GP is indicated by $\mathcal{GP}\left(m(\cdot),k(\cdot,\cdot)\right)$, where $m(\cdot)$ and $k(\cdot,\cdot)$ are the mean and kernel/covariance functions, respectively. For a sequence of matrices $X_1,\dots,X_n$, we denote the row-wise and diagonal-wise concatenation by $\col{X_1,\dots,X_n}$ and $\blkdiag{X_1,\dots,X_n}$, respectively. For a vector $x$, $\norm{x}_P$ denotes the weighted $l_2$-norm $(x^\top Px)^{\frac{1}{2}}$. The trace, determinant, log-determinant, and column vector containing the diagonal elements of a square matrix are indicated by $\tr{\cdot}$, $\detnew{\cdot}$, $\logdet{\cdot}$, and $\diag{\cdot}$, respectively. The vectorization, null space, and the $(i,j)$-th element of a matrix are denoted by $\vect{\cdot}$, $\nullsp{\cdot}$, and $(\cdot)_{i,j}$, respectively. Given a signal $x:\left\{1,\dots,N\right\}\to \mathbb{R}^n$, the Hankel matrix of depth $L$ is defined as
\begin{equation*}
    \mathcal{H}_L(\mathbf{x})=\begin{bmatrix}
        x_1&x_2&\cdots&x_{N-L+1}\\
        x_2&x_3&\cdots&x_{N-L+2}\\
        \vdots&\vdots&\ddots&\vdots\\
        x_L&x_{L+1}&\cdots&x_{N}
    \end{bmatrix},
\end{equation*}
and its trajectory from $i$ to $j$ is indicated by $(x_k)_{k=i}^{j}=\col{x_i,\dots,x_j}$. The symbols $\otimes$ and $^\dagger$ denote the Kronecker product and the Moore–Penrose inverse, respectively.

\begin{figure}[ht]
\center
\tikzstyle{block} = [draw,rectangle,thick,minimum height=2em,minimum width=1.25cm]%
\tikzstyle{sum} = [draw,circle,inner sep=0mm,minimum size=2mm]%
\tikzstyle{connector} = [->,thick]%
\tikzstyle{line} = [thick]%
\tikzstyle{branch} = [circle,inner sep=0pt,minimum size=1mm,fill=black,draw=black]%
\tikzstyle{guide} = []%

\tikzstyle{deltablock} = [block]%
\tikzstyle{controlblock} = [block]%
\tikzstyle{weightblock} = [block]%
\tikzstyle{clpblock} = [block]%

\begin{tikzpicture}[scale=1, auto, >=stealth']
    \small
    \node[block] (G) {$G(q)$};
    \node[block] (U) at ($(G.east) + (1.5cm,0)$) {$\psi(\cdot)$}; 
    \node[block] (Y) at ($(G.west) - (2.4cm,0)$) {$\phi^{-1}(\cdot)$}; 
    \node[sum] (esum) at ($(G.west) - (0.9cm,0)$) {+}; 
    
    \draw[connector] ($(U) + (1.5cm,0)$) -- node [midway, above] {$u_k$} (U);
    \draw[connector] (U) -- node [midway, above] {$\bar{u}_k$} (G);
    \draw[connector] (G) -- node [midway, above] {$\bar{y}_{k,0}$} (esum);
    \draw[connector] ($(esum) + (0,1cm)$) -- node [pos=0, left] {$e_k$} (esum);
    \draw[connector] (esum) -- node [midway, above] {$\bar{y}_k$} (Y);
    \draw[connector] (Y) -- node [midway, above] {$y_k$} ($(Y) - (1.5cm,0)$);
\end{tikzpicture}
\caption{Block diagram of Hammerstein-Wiener systems.}
\label{fig:hwsys}
\end{figure}
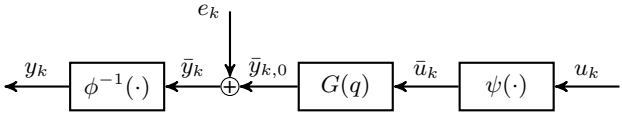

\section{Problem Statement and Background}
\label{sec:2}

Consider a H-W system shown in Fig.~\ref{fig:hwsys}, given by
\begin{equation}
\begin{cases}
x_{k+1}&=\ A x_k+B \bar{u}_k,\qquad\,\quad \bar{u}_k = \psi(u_k),\\
\hfil \bar{y}_k&=\ C x_k + D \bar{u}_k + e_k,\quad y_k = \phi^{-1}(\bar{y}_k),
\label{eq:sys}
\end{cases}
\end{equation}
where $x_k \in \mathbb{R}^{n_x}$, $u_k \in \mathbb{R}^{n_u}$, $y_k \in \mathbb{R}^{n_y}$ are the states, inputs, and outputs, respectively. The linear part of the system is denoted by $G(q)$, where $\bar{u}_k\in\mathbb{R}^{n_u}$, $\bar{y}_{k,0}\in\mathbb{R}^{n_y}$, $\bar{y}_k\in\mathbb{R}^{n_y}$, $e_k \in \mathbb{R}^{n_y}$ are the inputs, noise-free outputs, outputs, and output noise of $G(q)$, respectively. We also define $y_{k,0} = \phi^{-1}(\bar{y}_{k,0})$ as the noise-free nonlinear output. The functions $\psi(\cdot):\mathcal{U}\rightarrow\bar{\mathcal{U}}$, $\phi^{-1}(\cdot):\bar{\mathcal{Y}}\rightarrow\mathcal{Y}$, where $\mathcal{U},\bar{\mathcal{U}}\subseteq\mathbb{R}^{n_u}$ and $\mathcal{Y},\bar{\mathcal{Y}}\subseteq\mathbb{R}^{n_y}$ denote the input and output nonlinearities, respectively.

In this paper, the following assumption is considered.
\begin{assumption}
    1) The output noise $e_k \in \mathbb{R}^{n_y}$ is zero-mean i.i.d. Gaussian with covariance $\sigma^2\mathbb{I}$, 2) the linear part of the system $G(q)$ is stable, and 3) $\phi(\cdot)$ is monotonically increasing, i.e., the Jacobian of $\phi(y)$ is positive-definite for all $y$.
    \label{ass:1}
\end{assumption}
Assumption~\ref{ass:1}.3 guarantees that $\phi^{-1}(\cdot)$ is well-defined and is reasonable when $\phi^{-1}(\cdot)$ depicts, for example, sensor nonlinearity.

In this work, we are interested in characterizing probable length-$L$ input-output trajectories of \eqref{eq:sys} from a collected input-output trajectory  $\left(\mathbf{u}^d,\mathbf{y}^d\right):=\left(u_k^d,y_k^d\right)_{k=1}^N$ by incorporating the H-W model structure but not model knowledge of the components $\psi(\cdot)$, $G(q)$, and $\phi^{-1}(\cdot)$. This characterization is then used to design a receding horizon control algorithm.

\subsection{Data-Driven Prediction for Linear Systems}

This problem has been widely studied for linear systems, i.e., when $\psi(\cdot)$ and $\phi^{-1}(\cdot)$ are known. One successful approach recently relies on WFL \cite{Willems_2005}, skipping the explicit step of system identification. In detail, for $\mathbf{u}=(u_k)_{k=1}^L$ and $\mathbf{y}=(y_k)_{k=1}^L$, define $\Psi(\mathbf{u}):=\col{\psi(u_1),\dots,\psi(u_L)}$ and $\Phi(\mathbf{y}):=\col{\phi(y_1),\dots,\phi(y_L)}$. If noise-free trajectory data are available, we have the following lemma, which is a direct application of WFL, originally proposed in \cite[Theorem 1]{Willems_2005}, on the linear part of the H-W system.
\begin{lemma}
    Suppose $\left(\mathbf{u}^d,\mathbf{y}^d\right):=\left(u_k^d,y_k^d\right)_{k=1}^N$ is a noise-free trajectory of the H-W system \eqref{eq:sys}. Let $\bar{\mathbf{u}}^d := \Psi(\mathbf{u}^d)$, $\bar{\mathbf{y}}^d := \Phi(\mathbf{y}^d)$. If $\bar{\mathbf{u}}^d$ is persistently exciting of order $(L+n_x)$, i.e., $\mathcal{H}_{L+n_x}(\bar{\mathbf{u}}^d)$ has full row rank, and $G(q)$ is controllable, an input-output sequence $(\mathbf{u},\mathbf{y}):=\left(u_k,y_k\right)_{k=1}^L$ is a trajectory of \eqref{eq:sys} iff there exists $g\in\mathbb{R}^{M}$, $M=N-L+1$, such that
    \begin{equation*}
        \begin{bmatrix}
            H_{\bar{u}}\\
            H_{\bar{y}}
        \end{bmatrix}g=
        \begin{bmatrix}
            \Psi(\mathbf{u})\\
            \Phi(\mathbf{y}) - \mathbf{e}
        \end{bmatrix},
    \end{equation*}
    where $H_{\bar{u}}:=\mathcal{H}_L(\bar{\mathbf{u}}^d)$, $H_{\bar{y}}:=\mathcal{H}_L(\bar{\mathbf{y}}^d)$, and $\mathbf{e}:=\left(e_k\right)_{k=1}^L$.
    \label{lm:1}
\end{lemma}

Lemma~\ref{lm:1} is not directly applicable to our problem since we have neither noise-free trajectory data, nor the knowledge of the nonlinearities. Nevertheless, it is instructive for providing a suitable structure to learn the predictor, as detailed below. 

Let $L =: L_0+L'$, where $L_0$ is no smaller than the lag of $G(q)$. Define the following partitions:
\begin{equation*}
    \mathbf{y} =: \begin{bmatrix}
        \mathbf{y}_\mathrm{p}\\\mathbf{y}_\mathrm{f}
    \end{bmatrix},\ 
    \mathbf{e} =: \begin{bmatrix}
        \mathbf{e}_\mathrm{p}\\\mathbf{e}_\mathrm{f}
    \end{bmatrix},\ 
    H_{\bar{y}} =: \begin{bmatrix}
        H_{\bar{y}\mathrm{p}}\\H_{\bar{y}\mathrm{f}}
    \end{bmatrix},
\end{equation*}
where $\mathbf{y}_\mathrm{p},\mathbf{e}_\mathrm{p}\in\mathbb{R}^{n_y L_0}$, $\mathbf{y}_\mathrm{f},\mathbf{e}_\mathrm{f}\in\mathbb{R}^{n_y L'}$, $H_{\bar{y}\mathrm{p}}\in\mathbb{R}^{n_y L_0\times M}$, and $H_{\bar{y}\mathrm{f}}\in\mathbb{R}^{n_y L'\times M}$. Then, there exists a mapping from $\Psi(\mathbf{u})$ and $\Phi(\mathbf{y}_\mathrm{p})$ to $\Phi(\mathbf{y}_\mathrm{f})$, which is characterized by the following corollary.
\begin{corollary}
    Under the same conditions as Lemma~\ref{lm:1}, an input-output sequence $(\mathbf{u},\mathbf{y})$ is a trajectory of \eqref{eq:sys} iff it satisfies
    \begin{equation}
        \mathbf{0} = \left[\Gamma_1\ \Gamma_2\ -\mathbb{I}\right]\col{\Psi(\mathbf{u}),\Phi(\mathbf{y}_\mathrm{p}),\Phi(\mathbf{y}_\mathrm{f})}-\left[\Gamma_2\ -\mathbb{I}\right]\mathbf{e}.
        \label{eq:arx}
    \end{equation}
    where
    \begin{equation}
        \left[\Gamma_1\ \Gamma_2\right]:=H_{\bar{y}\mathrm{f}}\col{H_{\bar{u}},H_{\bar{y}\mathrm{p}}}^\dagger\in\mathbb{R}^{n_yL'\times (n_uL+n_yL_0)}.
        \label{eq:gam12}
    \end{equation}
    \label{cor:1}
\end{corollary}
\begin{pf}
    First note that $(\mathbf{u},\mathbf{y})$ is a trajectory of \eqref{eq:sys} iff $\left(\Psi(\mathbf{u}),\Phi(\mathbf{y}) - \mathbf{e}\right)$ is a trajectory of $G(q)$.
    According to \cite[Proposition 1]{Markovsky_20080}, when $L_0$ is no smaller than the lag of $G(q)$, $\left(\Psi(\mathbf{u}),\Phi(\mathbf{y}) - \mathbf{e}\right)$ is a trajectory of $G(q)$ iff $\Phi(\mathbf{y}_\mathrm{f}) - \mathbf{e}_\mathrm{f}=H_{\bar{y}\mathrm{f}}g$, where is $g$ is any solution to
    \begin{equation}
        \begin{bmatrix}
            H_{\bar{u}}\\
            H_{\bar{y}\mathrm{p}}
        \end{bmatrix}g=
        \begin{bmatrix}
            \Psi(\mathbf{u})\\
            \Phi(\mathbf{y}_\mathrm{p}) - \mathbf{e}_\mathrm{p}
        \end{bmatrix}.
        \label{eq:cor1}
    \end{equation}
    The solution to \eqref{eq:cor1} can be expressed as
    $$g = \col{H_{\bar{u}},H_{\bar{y}\mathrm{p}}}^\dagger\col{\Psi(\mathbf{u}),\Phi(\mathbf{y}_\mathrm{p}) - \mathbf{e}_\mathrm{p}}+g_0,$$
    where $g_0\in\nullsp{\col{H_{\bar{u}},H_{\bar{y}\mathrm{p}}}}$. Note that $H_{\bar{y}\mathrm{f}}g_0=\mathbf{0}$ from the proof of \cite[Theorem~1]{Fiedler_2021}. This leads to
    \begin{equation*}
        \Phi(\mathbf{y}_\mathrm{f}) - \mathbf{e}_\mathrm{f}=H_{\bar{y}\mathrm{f}}\col{H_{\bar{u}},H_{\bar{y}\mathrm{p}}}^\dagger\col{\Psi(\mathbf{u}),\Phi(\mathbf{y}_\mathrm{p}) - \mathbf{e}_\mathrm{p}},
    \end{equation*}
    which is equivalent to \eqref{eq:arx}.\qed
\end{pf}
For the rest of the paper, we define $\bar{\Gamma}_2:=\left[\Gamma_2\ -\mathbb{I}\right]$ for notational simplicity.
\begin{remark}
    Equation \eqref{eq:arx} can be seen as a multi-step-ahead nonlinear ARX model:
    \begin{equation}
        \mathbf{y}_\mathrm{f} = \Phi^{-1}\left(\left[\Gamma_1\ \Gamma_2\right]\col{\Psi(\mathbf{u}),\Phi(\mathbf{y}_\mathrm{p})}-\bar{\Gamma}_2\mathbf{e}\right).
        \label{eq:narx}
    \end{equation}
    For linear systems, this predictor is known as the subspace predictor \cite{Favoreel_1999}, and is equivalent to the solution to the least-squares problem in the noise-free case \cite{Fiedler_2021}:
    $$\underset{\Gamma_1,\Gamma_2}{\mathrm{min}}\ \norm{H_{\bar{y}\mathrm{f}}-\left[\Gamma_1\ \Gamma_2\right]\col{H_{\bar{u}},H_{\bar{y}\mathrm{p}}}}_F^2.$$
\end{remark}
If the linear part of the model is known, $\Gamma_1$ and $\Gamma_2$ can alternatively be derived from state-space parameters \cite[Equation 14]{Iannelli_2021}:
\begin{equation}
    \Gamma_1 =: \left[\Gamma_{11}\ \Gamma_{12}\right],\ \Gamma_{11} = \mathcal{O}_{L'}\mathcal{P},\ \Gamma_{12} = \mathcal{T}_{L'},\ \Gamma_2 = \mathcal{O}_{L'}\mathcal{Q},
    \label{eq:gam_mdl}
\end{equation}
where
\begin{align*}
    \mathcal{P}&:=\mathcal{C}_{L_0}-A^{L_0}\mathcal{O}_{L_0}^\dagger\mathcal{T}_{L_0},\quad \mathcal{Q}:=A^{L_0}\mathcal{O}_{L_0}^\dagger,\\
    \mathcal{T}_i&:=\begin{bmatrix}
        D & & &\\
        CB&D& &\\
        \vdots&\vdots&\ddots&\\
        CA^{i-2}B&CA^{i-3}B&\dots&D
    \end{bmatrix},\ \mathcal{O}_i:=\begin{bmatrix}
        C\\CA\\\vdots\\CA^{i-1}
    \end{bmatrix},\\
    \mathcal{C}_i&:=\begin{bmatrix}
        A^{i-1}B&\cdots&AB&B
    \end{bmatrix}.
\end{align*}
This model-based characterization provides insights into hyperprior design later in Section~\ref{sec:3c}.

\begin{remark}
    This work considers an output error structure for $G(q)$ with measurement noise. Alternatively, one can consider an ARX structure with equation errors for $G(q)$, i.e., $\bar{y}_k=\beta_0^u\bar{u}_k+\sum_{i=1}^{n_x}(\beta_i^u\bar{u}_{k-i}+\beta_i^y\bar{y}_{k-i})+e_k$. This leads to a different noise model $W^{-1}\mathbf{e}_\mathrm{f}$ \cite[Proposition~1]{Chiuso_2025}, where
    \begin{equation*}
        W = \begin{bmatrix}
            1&0&0&\cdots&0\\
            -\beta_1^y&1&0&\cdots&0\\
            -\beta_2^y&-\beta_1^y&1&\cdots&0\\
            \vdots&\vdots&\vdots&\ddots&\vdots\\
            -\beta_{L'-1}^y&-\beta_{L'-2}^y&-\beta_{L'-3}^y&\cdots&1\\
        \end{bmatrix},
    \end{equation*}
    in place of $-\bar{\Gamma}_2\mathbf{e}$ in \eqref{eq:arx}. A similar GP regression procedure as in Section~\ref{sec:3} can be applied, but with an additional hyperparameter $W$.
\end{remark}

\subsection{From Basis Functions to Implicit Gaussian Processes}
\label{sec:2b}

Based on the linear results in the previous subsection, an immediate strategy to analyze H-W systems with unknown input and output nonlinearities is to assume that finite-dimensional basis functions are available for the nonlinearities \cite{Berberich_2020}, i.e.,
\begin{align*}
    \psi(\cdot)&=:\sum_{i=1}^r a_i \psi_i(\cdot)=:\left(\mathbf{a}^\top\otimes\mathbb{I}\right)\boldsymbol{\psi}(\cdot),\\
    \phi(\cdot)&=:\sum_{i=1}^q b_i \phi_i(\cdot)=:\left(\mathbf{b}^\top\otimes\mathbb{I}\right)\boldsymbol{\phi}(\cdot),
\end{align*}
or equivalently $\psi(\cdot)\in\mathcal{H}_\psi$, $\phi(\cdot)\in\mathcal{H}_\phi$, where $\mathcal{H}_\psi$, $\mathcal{H}_\phi$ are the RKHS's associated to $k_\psi(u,u'):=\boldsymbol{\psi}(u)^\top\boldsymbol{\psi}(u')$, $k_\phi(y,y'):=\boldsymbol{\phi}(y)^\top\boldsymbol{\phi}(y')$, respectively. Considering the nonlinear system in the lifted input-output space leads to 
\begin{equation}
\begin{cases}
\hfil x_{k+1}&=\ A x_k+\bar{B}\boldsymbol{\psi}(u_k),\\
\left(\mathbf{b}^\top\otimes\mathbb{I}\right)\boldsymbol{\phi}(y_k)&=\ C x_k + \bar{D} \boldsymbol{\psi}(u_k),
\label{eq:syslin}
\end{cases}
\end{equation}
where $\bar{B}:=B \left(\mathbf{a}^\top\otimes\mathbb{I}\right)$ and $\bar{D}:=D \left(\mathbf{a}^\top\otimes\mathbb{I}\right)$. By considering augmented inputs $\boldsymbol{\psi}(u_k)$ and outputs $\boldsymbol{\phi}(y_k)$, \eqref{eq:syslin} admits linearity in the sense that for any two input-output trajectories of \eqref{eq:syslin}, their linear combinations are still valid trajectories. However, as also pointed out in \cite{Berberich_2020}, \eqref{eq:syslin} is not a standard linear model since the operator $\left(\mathbf{b}^\top\otimes\mathbb{I}\right)\in\mathbb{R}^{n_y\times n_yq}$ is not invertible. This prohibits the use of WFL to quantify all possible input-output trajectories and provide data-driven predictions when unknown nonlinear output uncertainties exist. Kernelized reformulations of WFL are presented in \cite{Molodchyk_2024,pmlr-v144-lian21a}. However, similar issues restrict its application to Wiener systems.

Even when we restrict ourselves to Hammerstein systems, a finite-dimensional basis function decomposition is often difficult to obtain, and the application of WFL requires persistency of excitation in the lifted signal space, which becomes difficult as the basis function dimension increases \cite{Alsalti_2023_2}. Although an infinite-dimensional function space associated to a general kernel can be applied with the kernelized reformulation, the persistency of excitation condition can no longer be satisfied, and exact trajectory certification becomes impossible \cite{Huang_2024}.

Since a straightforward application of WFL is difficult, an alternative strategy is to directly learn the nonlinear ARX model \eqref{eq:narx} for H-W systems. Yet, to the best of our knowledge, existing function learning methods cannot encode the composite nonlinear structure in \eqref{eq:narx}. One can, of course, disregard the specific structure in \eqref{eq:narx} and learn a purely black-box model $\mathbf{y}_\mathrm{f} = f(\mathbf{u},\mathbf{y}_\mathrm{p})$ by e.g., basis function decomposition \cite{Lazar_2024}, kernel regression \cite{Huang_2024}, or GP regression \cite{1383790}. However, this would lead to deteriorated performance since a much larger function space is learned from, and the predictor would not correspond to an H-W system.

Due to the above reasons, in this work, we employ the implicit trajectory characterization \eqref{eq:arx} instead of the explicit one \eqref{eq:narx}, since \eqref{eq:arx} does not contain any composite of nonlinear functions. This enables a physics-informed design that incorporates the H-W structure in learning. Implicit GP regression is applied to jointly learn the linear mappings $\Gamma_1$, $\Gamma_2$ and the nonlinearities $\psi(\cdot)$, $\phi(\cdot)$, as will be detailed in the following section.

\section{Implicit Gaussian Process Data-Driven Prediction}
\label{sec:3}

This section presents the implicit GP data-driven prediction scheme for H-W systems. Section~\ref{sec:3a} starts by explaining the standard procedure of GP regression, where the structured kernel design for the implicit GP is obtained in Section~\ref{sec:3b}. Section~\ref{sec:3b1} incorporates monotonicity information in the regression by expectation propagation. The posterior predictions from the implicit model are given in Section~\ref{sec:3b2}. Section~\ref{sec:3c} investigates the hyperparameter tuning problem for the linear model parameters, followed by Section~\ref{sec:3d}, which presents a scheme to recover the nonlinear functions.

\subsection{Gaussian Process Regression}
\label{sec:3a}

The standard procedure of GP regression is briefly summarized as follows \cite{10.7551/mitpress/3206.001.0001}. Consider the problem of learning an unknown function $f:\mathbb{R}^{n_{\eta}}\rightarrow\mathbb{R}^{n_\chi}$ from data $\left(\boldsymbol{\boldsymbol{\eta}}^d,\boldsymbol{\chi}^d\right)=\left(\eta_k^d,\chi_k^d\right)_{k=1}^M$, where $\chi^d_k=f(\eta^d_k)+\epsilon_k$ and $\epsilon_k$ is zero-mean i.i.d. Gaussian with covariance $\Sigma$. The function $f$ is modeled as a GP with prior mean $m(\cdot)$, prior covariance $k(\cdot,\cdot;\Theta)$, and hyperparameters $\Theta$, i.e.,
$f(\cdot)\sim\mathcal{GP}\left(m(\cdot),k(\cdot,\cdot;\Theta)\right)$. To predict $f(\eta)$ at any query point $\eta$, the joint distribution of $f(\eta)$ and $\boldsymbol{\chi}^d$ is given by the following Gaussian distribution:
\begin{equation*}
    \begin{bmatrix}
        f(\eta)\\\boldsymbol{\chi}^d
    \end{bmatrix}\sim\mathcal{N}\left(\begin{bmatrix}
        m(\eta)\\\mathbf{m}(\boldsymbol{\eta}^d)
    \end{bmatrix},\begin{bmatrix}
        k(\eta,\eta)&\mathbf{k}^\top(\boldsymbol{\eta}^d,\eta)\\
        \mathbf{k}(\boldsymbol{\eta}^d,\eta)&K(\boldsymbol{\eta}^d,\boldsymbol{\eta}^d)+\bar{\Sigma}
    \end{bmatrix}
    \right),
\end{equation*}
where $\bar{\Sigma}:=\mathbb{I}\otimes\Sigma$,
\begin{align}
\mathbf{m}(\boldsymbol{\eta}^d)&:=\col{m(\eta^d_1),\dots,m(\eta^d_M)},\label{eq:vecm}\\
\mathbf{k}(\boldsymbol{\eta}^d,\eta)&:=\col{k(\eta^d_1,\eta),\dots,k(\eta^d_M,\eta)},\\
K(\boldsymbol{\eta}^d,\boldsymbol{\eta}^d)&:=\begin{bmatrix}
    k(\eta^d_1,\eta^d_1)&k(\eta^d_1,\eta^d_2)&\cdots&k(\eta^d_1,\eta^d_M)\\
    k(\eta^d_2,\eta^d_1)&k(\eta^d_2,\eta^d_2)&\cdots&k(\eta^d_2,\eta^d_M)\\
    \vdots&\vdots&\ddots&\vdots\\
    k(\eta^d_M,\eta^d_1)&k(\eta^d_M,\eta^d_2)&\cdots&k(\eta^d_M,\eta^d_M)\label{eq:matk}\\
\end{bmatrix}.
\end{align}
This leads to the conditional probability of $f(\eta)|\boldsymbol{\chi}^d$ as
\begin{equation}
    f(\eta)|\boldsymbol{\chi}^d\sim\mathcal{N}\left(m_\mathrm{p}(\eta),k_\mathrm{p}(\eta)\right),
    \label{eq:distpost}
\end{equation}
where
\begin{align}
    m_\mathrm{p}(\eta)&:=m(\eta)+\mathbf{k}^\top(\boldsymbol{\eta}^d,\eta)\left(K+\bar{\Sigma}\right)^{-1}\left(\boldsymbol{\chi}^d-\mathbf{m}(\boldsymbol{\eta}^d)\right),\label{eq:mpost}\\
    k_\mathrm{p}(\eta)&:=k(\eta,\eta)-\mathbf{k}^\top(\boldsymbol{\eta}^d,\eta)\left(K+\bar{\Sigma}\right)^{-1}\mathbf{k}(\boldsymbol{\eta}^d,\eta),\label{eq:kpost}
\end{align}
where the dependence of $K$ on $\boldsymbol{\eta}^d$ is omitted for brevity. Then, $f(\eta)$ is predicted as the posterior mean $m_\mathrm{p}(\eta)$.

The prediction is conditioned on a particular value of the hyperparameters $\Theta$, which is often estimated by using the maximum marginal likelihood method, also known as the empirical Bayes method, which maximizes the probability of observing $\boldsymbol{\chi}^d$ given the observed trajectories $\boldsymbol{\eta}^d$ and the hyperparameters $\Theta$ by considering $\Theta$ as deterministic variables, i.e.,
\begin{align}
    \Theta &= \mathrm{arg}\underset{\Theta}{\mathrm{max}}\ p\left(\boldsymbol{\chi}^d|\Theta\right)\\
    &= \mathrm{arg}\underset{\Theta}{\mathrm{min}}\ \logdet{\Lambda(\Theta)}+\delta^\top \Lambda^{-1}(\Theta)\delta,\label{eq:mml}
\end{align}
where $\delta:=\boldsymbol{\chi}^d-\mathbf{m}(\boldsymbol{\eta}^d)$ and $\Lambda(\Theta):=K(\boldsymbol{\eta}^d,\boldsymbol{\eta}^d;\Theta)+\bar{\Sigma}$.

\subsection{Structured Kernel Design}
\label{sec:3b}

Since the input and output nonlinearities are unknown, they are modeled as GPs with prior distributions:
\begin{equation}
    \begin{aligned}
        \psi(\cdot)&\sim\mathcal{GP}\left(m_u(\cdot),k_u(\cdot,\cdot;\theta_u)\right),\\
        \phi(\cdot)&\sim\mathcal{GP}\left(m_y(\cdot),k_y(\cdot,\cdot;\theta_y)\right),
    \end{aligned}
    \label{eq:gpmdl}
\end{equation}
where $m_u:\mathbb{R}^{n_u}\rightarrow\mathbb{R}^{n_u}$, $m_y:\mathbb{R}^{n_y}\rightarrow\mathbb{R}^{n_y}$ are the mean functions,  $k_u:\mathbb{R}^{n_u}\times\mathbb{R}^{n_u}\rightarrow\mathbb{R}^{n_u}$, $k_y:\mathbb{R}^{n_y}\times\mathbb{R}^{n_y}\rightarrow\mathbb{R}^{n_y}$ are the kernel/covariance functions, and $\theta_u$ and $\theta_y$ are hyperparameters for the kernel.
\begin{remark}
    Most standard kernel structures include a scaling parameter as a hyperparameter. In our setting, the model \eqref{eq:sys} is ambiguous up to scaling factors, i.e., the model triplet $\left(\psi(\cdot),G(q),\phi^{-1}(\cdot)\right)$ is equivalent to $\left(a\psi(\cdot),\frac{1}{ab}G(q),b\phi^{-1}(\cdot)\right)$, $a,b\in\mathbb{R}$. Thus, the scaling parameter can be fixed to 1 without loss of generality. Similarly, the mean functions $m_u$ and $m_y$ can also be normalized.
    \label{rm:40}
\end{remark}
\begin{remark}
    When input or output nonlinearities are absent, an identity prior mean function and a zero kernel function can be selected. For example, for Hammerstein systems, we can set $m_y(y)=y$ and $k_y(\cdot,\cdot)=0$.
    \label{rm:4}
\end{remark}

Although the prior distribution of $\Gamma_1,\Gamma_2$ can be derived from \eqref{eq:gam12} and the Gaussian prior \eqref{eq:gpmdl}, it is dependent on $\psi(\cdot)$ and $\phi(\cdot)$ in a nonlinear way, and thus the Gaussianity is not preserved. Instead, we opt for directly learning $\Gamma_1,\Gamma_2$ in a hierarchical model by regarding them as hyperparameters, as will be discussed in Section~\ref{sec:3c}.

The implicit GP regression problem is formulated as follows. Consider the trajectory characterization equation \eqref{eq:arx}. Let $\eta=\col{\mathbf{u},\mathbf{y}}$ and
\begin{equation}
    f(\eta)=\left[\Gamma_1\ \bar{\Gamma}_2\right]\col{\Psi(\mathbf{u}),\Phi(\mathbf{y}_\mathrm{p}),\Phi(\mathbf{y}_\mathrm{f})},
    \label{eq:feta}
\end{equation}
where $n_\eta=(n_u+n_y)L$ and $n_\chi=n_yL'$. We would like to train a GP model of $f(\eta)$ by using data $\left(\boldsymbol{\boldsymbol{\eta}}^d,\boldsymbol{\chi}^d\right)$ where $M=N-L+1$,
\begin{align}
    \eta_k^d&=\col{u^d_k,\dots,u^d_{k+L-1},y^d_k,\dots,y^d_{k+L-1}},\chi_k^d=\mathbf{0},\label{eq:tdata}\\
    \bar{\Sigma}&=\mathbb{I}\otimes\cov{\bar{\Gamma}_2\mathbf{e}}=\sigma^2\left(\mathbb{I}\otimes\left(\Gamma_2\Gamma_2^\top+\mathbb{I}\right)\right).\label{eq:noisemdl}
\end{align}
This is known as an implicit GP model \cite{Martens_2017}.

The prior mean and covariance functions $m(\cdot)$ and $k(\cdot,\cdot)$ for this implicit GP model are then derived from the GP characterization of $\psi(\cdot)$ and $\phi(\cdot)$ in \eqref{eq:gpmdl}. The mean function of an input-output trajectory $\eta=\col{\mathbf{u},\mathbf{y}}$ is given by
\begin{equation}
    m(\eta)=\left[\Gamma_1\ \bar{\Gamma}_2\right]\col{\mathbf{m}_u(\mathbf{u}),\mathbf{m}_y(\mathbf{y})},
    \label{eq:m}
\end{equation}
where $\mathbf{m}_u(\cdot)$, $\mathbf{m}_y(\cdot)$ apply $m_u(\cdot)$, $m_y(\cdot)$ elementwise, respectively. Consider two input-output trajectories $\eta'=\col{\mathbf{u}',\mathbf{y}'}$ and $\eta''=\col{\mathbf{u}'',\mathbf{y}''}$. The covariance matrix between $\Psi(\mathbf{u}')$ and $\Psi(\mathbf{u}'')$ is given by $\cov{\Psi(\mathbf{u}'),\Psi(\mathbf{u}'')}=\mathbf{k}_u(\mathbf{u}',\mathbf{u}'')\in\mathbb{R}^{n_uL\times n_uL}$, and similarly $\cov{\Phi(\mathbf{y}'),\Phi(\mathbf{y}'')}=\mathbf{k}_y(\mathbf{y}',\mathbf{y}'')\in\mathbb{R}^{n_yL\times n_yL}$, where $\mathbf{k}_u(\cdot,\cdot)$, $\mathbf{k}_y(\cdot,\cdot)$ apply $k_u(\cdot,\cdot)$, $k_y(\cdot,\cdot)$ elementwise, respectively.

Then, we have $\cov{f(\eta'),f(\eta'')}=k(\eta',\eta'')$, where
\begin{equation}
    k(\eta',\eta''):=\Gamma_1\mathbf{k}_u(\mathbf{u}',\mathbf{u}'')\Gamma_1^\top+\bar{\Gamma}_2\mathbf{k}_y(\mathbf{y}',\mathbf{y}'')\bar{\Gamma}_2^\top.
    \label{eq:priork}
\end{equation}
To simplify calculation, substituting \eqref{eq:m} and \eqref{eq:priork} into \eqref{eq:vecm}-\eqref{eq:matk}, after some algebraic manipulation, we have
\begin{align}
\mathbf{m}(\boldsymbol{\eta}^d)&=\vect{\left[\Gamma_1\ \bar{\Gamma}_2\right]\begin{bmatrix}
    \mathbf{m}_u(H_u)\\\mathbf{m}_y(H_y)
\end{bmatrix}},\label{eq:mfinal}\\
\mathbf{k}(\boldsymbol{\eta}^d,\eta)&=\left(\mathbb{I}\otimes\Gamma_1\right)\mathbf{k}_u(\vect{H_u},\mathbf{u})\Gamma_1^\top\notag\\&\quad+\left(\mathbb{I}\otimes \bar{\Gamma}_2\right)\mathbf{k}_y(\vect{H_y},\mathbf{y})\bar{\Gamma}_2^\top,\label{eq:kfinal}\\
K(\boldsymbol{\eta}^d,\boldsymbol{\eta}^d)&=\left(\mathbb{I}\otimes\Gamma_1\right)\mathbf{k}_u(\vect{H_u},\vect{H_u})\left(\mathbb{I}\otimes\Gamma_1^\top\right)\notag\\&\ +\left(\mathbb{I}\otimes \bar{\Gamma}_2\right)\mathbf{k}_y(\vect{H_y},\vect{H_y})\left(\mathbb{I}\otimes \bar{\Gamma}_2^\top\right)\label{eq:Kfinal},
\end{align}
where $H_u:=\mathcal{H}_L(\mathbf{u}^d)$, $H_y:=\mathcal{H}_L(\mathbf{y}^d)$. The detailed calculation is presented in Appendix~\ref{app:1}.

By applying the structured kernel design \eqref{eq:mfinal}-\eqref{eq:Kfinal}, the GP model \eqref{eq:distpost}-\eqref{eq:kpost} of the implicit function $f(\eta)$ can be obtained with prior mean and covariance functions \eqref{eq:m}, \eqref{eq:priork}, noise model \eqref{eq:noisemdl}, and training data \eqref{eq:tdata}.

For later derivations, we also define the joint distribution of $\Phi(\mathbf{y})$ and $\boldsymbol{\chi}^d$:
\begin{equation}
\resizebox{\columnwidth}{!}{$
    \begin{bmatrix}
        \Phi(\mathbf{y})\\\boldsymbol{\chi}^d
    \end{bmatrix}\sim\mathcal{N}\left(\begin{bmatrix}
        \mathbf{m}_y(\mathbf{y})\\\mathbf{m}(\boldsymbol{\eta}^d)
    \end{bmatrix},\begin{bmatrix}
        \mathbf{k}_y(\mathbf{y},\mathbf{y})&\kappa_y^\top(\boldsymbol{\eta}^d,\mathbf{y})\\
        \kappa_y(\boldsymbol{\eta}^d,\mathbf{y})&K(\boldsymbol{\eta}^d,\boldsymbol{\eta}^d)+\bar{\Sigma}
    \end{bmatrix}
    \right),$}
    \label{eq:phichi}
\end{equation}
where, similar to the derivation in Appendix~\ref{app:1},
\begin{equation}
    \kappa_y(\boldsymbol{\eta}^d,\mathbf{y}):=\left(\mathbb{I}\otimes \bar{\Gamma}_2\right)\mathbf{k}_y(\vect{H_y},\mathbf{y}).
\end{equation}
Similarly, we also define the joint distribution of $\Psi(\mathbf{u})$ and $\boldsymbol{\chi}^d$ with
\begin{equation}
    \kappa_u(\boldsymbol{\eta}^d,\mathbf{u}):=\left(\mathbb{I}\otimes \Gamma_1\right)\mathbf{k}_u(\vect{H_u},\mathbf{u}).
\end{equation}

\subsection{Incorporating Monotonicity by Expectation Propagation}
\label{sec:3b1}

The GP model \eqref{eq:distpost}-\eqref{eq:kpost}, however, does not incorporate the monotonicity of the output nonlinearity $\phi(\cdot)$ as required in Assumption~\ref{ass:1}.3. Due to the unbounded nature of Gaussian processes, it is not possible to guarantee monotonicity for all posterior samples with certainty. Instead, since the derivative of a Gaussian process is still Gaussian, we follow the idea described in \cite{pmlr-v9-riihimaki10a} and add \textit{virtual derivative points} to the training data to promote functions that are monotonically increasing.

\begin{remark}
    We note that if the posterior mean estimate of $\phi(\cdot)$ (cf. Section~\ref{sec:3d}) is not monotonic, the predictor remains well-defined, but it does not correspond to a well-defined Wiener system.
\end{remark}

For simplicity of exposition, we present the case where $n_y=1$. The extension to the multi-output case is discussed in Remark~\ref{rm:33}. We would like to include information that $\phi'(y_{\mathrm{m},i})>0$ for a grid of virtual derivative points $\mathbf{y}_\mathrm{m}:=\left(y_{\mathrm{m},i}\right)_{i=1}^{n_m}$ in the GP regression scheme. See \cite[Section~3.1]{pmlr-v9-riihimaki10a} for alternative strategies to place virtual derivative points. Let $\mathcal{M}_i$ be the event that $\phi'(y_{\mathrm{m},i})>0$. Similar to the probit regression in classification \cite[Chapter~3]{10.7551/mitpress/3206.001.0001}, we use the following probit likelihood to approximate the monotonicity constraint $\phi'(y_{\mathrm{m},i})>0$ with a smooth function:
\begin{equation}
    \pr{\mathcal{M}_i|\phi'(y_{\mathrm{m},i})}\approx F_\mathcal{N}\left(\phi'(y_{\mathrm{m},i})/\nu_0\right),
    \label{eq:probit}
\end{equation}
where $F_\mathcal{N}(\cdot)$ is the cumulative distribution function of the unit Gaussian distribution and $\nu_0>0$ is a small constant such that $F_\mathcal{N}\left(x/\nu_0\right)$ approximates the step function as $\nu_0\rightarrow 0$. Following this approximation, we replace $\pr{\mathcal{M}_i|\phi'(y_{\mathrm{m},i})}$ by $F_\mathcal{N}\left(\phi'(y_{\mathrm{m},i})/\nu_0\right)$ in what follows.

Then, from Bayes' rule, we have
\begin{multline}
    p(\Phi'(\mathbf{y}_\mathrm{m}),\mathbf{f}(\boldsymbol{\eta}^d)|\mathcal{M},\boldsymbol{\chi}^d)\propto p(\Phi'(\mathbf{y}_\mathrm{m}),\mathbf{f}(\boldsymbol{\eta}^d))\\p(\boldsymbol{\chi}^d|\mathbf{f}(\boldsymbol{\eta}^d))\prod_{i=1}^{n_m}F_\mathcal{N}\left(\phi'(y_{\mathrm{m},i})/\nu_0\right),
    \label{eq:eppost}
\end{multline}
where $\Phi'(\mathbf{y}_\mathrm{m}):=\col{\phi'(y_{\mathrm{m},1}),\dots,\phi'(y_{\mathrm{m},n_m})}$, $\mathcal{M}:=\cap_{i=1}^{n_m}\mathcal{M}_i$, and $\boldsymbol{\chi}^d|\mathbf{f}(\boldsymbol{\eta}^d)\sim\mathcal{N}\left(\mathbf{f}(\boldsymbol{\eta}^d),\bar{\Sigma}\right)$. The joint distribution of $\mathbf{f}(\boldsymbol{\eta}^d)$ and $\Phi'(\mathbf{y}_\mathrm{m})$ is calculated as
\begin{equation}
\resizebox{\columnwidth}{!}{$
    \begin{bmatrix}
        \Phi'(\mathbf{y}_\mathrm{m})\\\mathbf{f}(\boldsymbol{\eta}^d)
    \end{bmatrix}\sim\mathcal{N}\left(
    \underbrace{
    \begin{bmatrix}
        \mathbf{m}'_y(\mathbf{y}_\mathrm{m})\\\mathbf{m}(\boldsymbol{\eta}^d)
    \end{bmatrix}}_{\mathbf{m}_\mathrm{j}},
    \underbrace{
    \begin{bmatrix}
        \mathbf{k}''_y(\mathbf{y}_\mathrm{m},\mathbf{y}_\mathrm{m})&{\kappa'_y}^\top(\boldsymbol{\eta}^d,\mathbf{y}_\mathrm{m})\\
        \kappa'_y(\boldsymbol{\eta}^d,\mathbf{y}_\mathrm{m})&K(\boldsymbol{\eta}^d,\boldsymbol{\eta}^d)
    \end{bmatrix}}_{K_\mathrm{j}}
    \right),$}
    \label{eq:monojoint}
\end{equation}
where $\mathbf{m}'_y(\mathbf{y}_\mathrm{m})$, $\mathbf{k}''_y(\mathbf{y}_\mathrm{m},\mathbf{y}_\mathrm{m})$, $\kappa'_y(\boldsymbol{\eta}^d,\mathbf{y}_\mathrm{m})$ apply $m'_y(y_{\mathrm{m},i})$, $\frac{\partial k_y(y_{\mathrm{m},i},y_{\mathrm{m},j})}{\partial y_{\mathrm{m},i}\partial y_{\mathrm{m},j}}$, $\frac{\partial \kappa_y(\boldsymbol{\eta}^d,y_{\mathrm{m},i})}{\partial y_{\mathrm{m},i}}$ elementwise, respectively.

Unfortunately, the posterior distribution \eqref{eq:eppost} is no longer Gaussian due to the 
terms $F_\mathcal{N}\left(\phi'(y_{\mathrm{m},i})/\nu_0\right)$. For tractability, we apply the expectation propagation (EP) algorithm \cite[Section~3.6]{10.7551/mitpress/3206.001.0001} by approximating $F_\mathcal{N}\left(\phi'(y_{\mathrm{m},i})/\nu_0\right)$ with an un-normalized Gaussian likelihood.

A numerically stable and efficient EP algorithm is presented in Algorithm~\ref{al:0}. This algorithm adapts \cite[Algorithm~3.5]{10.7551/mitpress/3206.001.0001} for classification problems to the virtual derivative point method, using the posterior marginal moment equations from \cite[Section~3]{pmlr-v9-riihimaki10a}. The details of the EP algorithm are discussed in Appendix~\ref{app:2}.

\begin{algorithm}[thb]
\caption{Expectation propagation for the virtual derivative point method}
\begin{algorithmic}[1]
    \State \textbf{Input:} $\mathbf{m}_\mathrm{j}$, $K_\mathrm{j}$, $\bar{\Sigma}$
    \State \textbf{Initialize:} $\mathbf{S}:=\left(K_\mathrm{j}^{-1}+\blkdiag{\mathbf{0},\bar{\Sigma}^{-1}}\right)^{-1}$, $\boldsymbol{\mu}:=\mathbf{S}K_\mathrm{j}^{-1}\mathbf{m}_\mathrm{j}$, $\boldsymbol{\nu}:=\mathbf{0}$, $\boldsymbol{\tau}:=\mathbf{0}$
    \Repeat
    \For{$i:=1$ to $n_m$}
    \State $\tau_{-i}:=(\mathbf{S})_{i,i}^{-1}-\tau_i$, $\nu_{-i}:=(\mathbf{S})_{i,i}^{-1}\mu_i-\nu_i$
    \State $z_i:=\dfrac{\nu_{-i}}{\nu_0\tau_{-i}\sqrt{1+1/(\nu_0^2\tau_{-i})}}$
    \State $\hat{\mu}_i:=\dfrac{\nu_{-i}}{\tau_{-i}}+\dfrac{p_{\mathcal{N}}(z_i)}{\nu_0\tau_{-i}F_\mathcal{N}(z_i)\sqrt{1+1/(\nu_0^2\tau_{-i})}}$
    \State 
    \vspace{-1.5em}
    $$\hat{\sigma}_i^2:=\dfrac{1}{\tau_{-i}}-\dfrac{p_{\mathcal{N}}(z_i)}{\tau_{-i}^2F_\mathcal{N}(z_i)(\nu_0^2+1/\tau_{-i})}\left(z_i+\dfrac{p_{\mathcal{N}}(z_i)}{F_\mathcal{N}(z_i)}\right)$$
    \vspace{-1em}
    \State $\Delta\tau:=\hat{\sigma}_i^{-2}-\tau_{-i}-\tau_i$
    \State $\tau_i:=\tau_i+\Delta\tau$, $\nu_i:=\hat{\sigma}_i^{-2}\hat{\mu}_i-\nu_{-i}$
    \State $\mathbf{S}:=\mathbf{S}-\left((\Delta\tau)^{-1}+(\mathbf{S})_{i,i}\right)^{-1}\mathbf{s}_i\mathbf{s}_i^\top$, where $\mathbf{s}_i$ is the $i$-th column of $\mathbf{S}$.
    \State $\boldsymbol{\mu} := \mathbf{S}\col{\boldsymbol{\nu},\mathbf{0}} + \mathbf{S}K_\mathrm{j}^{-1}\mathbf{m}_\mathrm{j}$
    \EndFor
    \Until{convergence}
    \State \textbf{Output:} $\tilde{\boldsymbol{\mu}}_\mathrm{j}:=\col{\nu_1/\tau_1,\dots,\nu_{n_m}/\tau_{n_m},\mathbf{0}}$, $\tilde{\Sigma}_\mathrm{j}:=\blkdiag{1/\tau_1,\dots,1/\tau_{n_m},\bar{\Sigma}}$.
\end{algorithmic}
\label{al:0}
\end{algorithm}

\begin{remark}
    When $n_y>1$, the monotonicity constraint is replaced by $\mathbf{J}_\phi(y_{\mathrm{m},i})\succ 0$, where $\mathbf{J}_\phi(\cdot)$ denotes the Jacobian of $\phi(\cdot)$. The probit likelihood \eqref{eq:probit} is thus replaced by
    \begin{equation}
        \pr{\mathcal{M}_i|\mathbf{J}_\phi(y_{\mathrm{m},i})}\approx F_\mathcal{N}\left(\detnew{\mathbf{J}_\phi(y_{\mathrm{m},i})}/\nu_0\right).
    \end{equation}
    The virtual derivative points become the vectorized Jacobian:
    $$\Phi'(\mathbf{y}_\mathrm{m})=\col{\vect{\mathbf{J}_\phi(y_{\mathrm{m},1})},\dots,\vect{\mathbf{J}_\phi(y_{\mathrm{m},n_m})}},$$
    and the definitions of $\mathbf{m}'_y(\mathbf{y}_\mathrm{m})$, $\mathbf{k}''_y(\mathbf{y}_\mathrm{m},\mathbf{y}_\mathrm{m})$, $\kappa'_y(\boldsymbol{\eta}^d,\mathbf{y}_\mathrm{m})$ are modified accordingly. The EP algorithm remains unchanged except that 1) the approximate Gaussian likelihood becomes multivariate of dimension $n_y^2$, and 2) the closed-form solutions to the desired posterior marginal moments $\hat{\mu}_i$, $\hat{\sigma}_i^2$ (lines 6-8 in Algorithm~\ref{al:0}) are no longer valid, cf. \cite[Section~3.9]{10.7551/mitpress/3206.001.0001}, and numerical integration is required to calculate the moments.
    \label{rm:33}
\end{remark}

From \eqref{eq:b6} and \eqref{eq:b7}, the approximate posterior distribution with the virtual derivative points is given by:
\begin{equation}
    f(\eta)|\mathcal{M},\boldsymbol{\chi}^d\sim\mathcal{N}\left(m_\mathrm{pm}(\eta),k_\mathrm{pm}(\eta)\right),
    \label{eq:distpostm}
\end{equation}
where
\begin{align}
    m_\mathrm{pm}(\eta)&:=m(\eta)+\mathbf{k}^\top_\mathrm{j}(\eta)(K_\mathrm{j}+\tilde{\Sigma}_\mathrm{j})^{-1}\left(\tilde{\boldsymbol{\mu}}_\mathrm{j}-\mathbf{m}_\mathrm{j}\right),\label{eq:mpostm}\\
    k_\mathrm{pm}(\eta)&:=k(\eta,\eta)-\mathbf{k}^\top_\mathrm{j}(\eta)(K_\mathrm{j}+\tilde{\Sigma}_\mathrm{j})^{-1}\mathbf{k}_\mathrm{j}(\eta),\label{eq:kpostm}
\end{align}
Similar to \eqref{eq:mml}, the hyperparameters can be estimated by maximizing the approximate marginal likelihood:
\begin{multline}
    q\left(\mathcal{M},\boldsymbol{\chi}^d|\Theta\right):=Z=(2\pi)^{-\frac{M+n_m}{2}}\detnew{\Lambda_\mathrm{j}(\Theta)}^{-\frac{1}{2}}\\\exp\left(-\frac{1}{2}\delta_\mathrm{j}^\top \Lambda_\mathrm{j}^{-1}(\Theta)\delta_\mathrm{j}\right)
    \prod_{i=1}^{n_m}\tilde{Z}_i,
    \label{eq:39}
\end{multline}
where we use $q(\cdot)$ to denote approximate probability densities after expectation propagation, $\delta_\mathrm{j}:=\tilde{\boldsymbol{\mu}}_\mathrm{j}-\mathbf{m}_\mathrm{j}$, and $\Lambda_\mathrm{j}(\Theta):=K_\mathrm{j}+\tilde{\Sigma}_\mathrm{j}$.
Luckily, it was shown in \cite[Appendix~C.1.2]{item_59ced79659794bafbf4055cb513607e9} that the gradient of $\tilde{Z}_i$ with respect to the hyperparameters is zero. Thus, the hyperparameter estimation problem can be simplified as
\begin{align}
    \Theta_\mathrm{m} &= \mathrm{arg}\underset{\Theta}{\mathrm{max}}\ q\left(\mathcal{M},\boldsymbol{\chi}^d|\Theta\right)\\
    &= \mathrm{arg}\underset{\Theta}{\mathrm{min}}\ \logdet{\Lambda_\mathrm{j}(\Theta)}+\delta_\mathrm{j}^\top \Lambda_\mathrm{j}^{-1}(\Theta)\delta_\mathrm{j}.\label{eq:mmlm}
\end{align}
\begin{remark}
    The approach presented in this section generalizes the results in \cite{pmlr-v9-riihimaki10a} by allowing for non-zero prior mean functions and virtual derivative points on components of the nonlinear function rather than the whole function. 
\end{remark}

\subsection{Posterior Prediction}
\label{sec:3b2}

From \eqref{eq:feta}, we have the posterior distribution
$$\left[\Gamma_1\ \Gamma_2\right]\col{\Psi(\mathbf{u}),\Phi(\mathbf{y}_\mathrm{p})}-\Phi(\mathbf{y}_\mathrm{f})\sim\mathcal{N}\left(m_\mathrm{pm}(\eta),k_\mathrm{pm}(\eta)\right).$$
Let the true noise-free future output trajectory be $\mathbf{y}_\mathrm{f,0}$. From \eqref{eq:arx}, we have $\left[\Gamma_1\ \Gamma_2\right]\col{\Psi(\mathbf{u}),\Phi(\mathbf{y}_\mathrm{p})}-\Phi(\mathbf{y}_\mathrm{f,0})=\left[\Gamma_2\ \mathbf{0}\right]\mathbf{e}$. Therefore, the GP model characterizes probable trajectories of \eqref{eq:sys} in the sense that for any given input trajectory $\mathbf{u}$ and past output trajectory $\mathbf{y}_\mathrm{p}$, $\mathbf{y}_\mathrm{f}$ is the future output trajectory with projected prediction error
\begin{equation}
    \Phi(\mathbf{y}_\mathrm{f,0})-\Phi(\mathbf{y}_\mathrm{f})\sim\mathcal{N}\left(m_\mathrm{pm}(\eta),\hat{k}_\mathrm{pm}(\eta)\right),
    \label{eq:pperror}
\end{equation}
where
\begin{equation}
    \hat{k}_\mathrm{pm}(\eta)=k_\mathrm{pm}(\eta)+\cov{\left[\Gamma_2\ \mathbf{0}\right]\mathbf{e}}=k_\mathrm{pm}(\eta)+\sigma^2\Gamma_2\Gamma_2^\top.
    \label{eq:khat}
\end{equation}
The optimal posterior prediction of $\mathbf{y}_\mathrm{f}$ can be obtained by the maximum likelihood method. In detail, $\mathbf{y}_\mathrm{f}$ is estimated by maximizing the probability of observing $f(\eta)=\mathbf{0}$, i.e.,
\begin{align}
    \mathbf{y}_\mathrm{f} &= \mathrm{arg}\underset{\mathbf{y}_\mathrm{f}}{\mathrm{max}}\ p\left(f(\eta)=\mathbf{0}|\mathcal{M},\boldsymbol{\chi}^d\right)\\
    &= \mathrm{arg}\underset{\mathbf{y}_\mathrm{f}}{\mathrm{min}}\ \logdet{k_\mathrm{pm}(\eta)}+m_\mathrm{pm}^\top(\eta)k_\mathrm{pm}^{-1}(\eta)m_\mathrm{pm}(\eta).\label{eq:predopt1}
\end{align}

\begin{remark}
Other optimality criteria can also be used to obtain the optimal prediction depending on user preferences. This includes the minimum mean-squared error predictor:
    \begin{equation}
        \mathbf{y}_\mathrm{f}=\mathrm{arg}\underset{\mathbf{y}_\mathrm{f}}{\mathrm{min}}\ \mathrm{tr}\left(\hat{k}_\mathrm{pm}(\eta)\right)+m_\mathrm{pm}^\top(\eta)m_\mathrm{pm}(\eta),\label{eq:predopt2}
    \end{equation}
and the minimum-variance unbiased predictor:
    \begin{equation}
        \mathbf{y}_\mathrm{f}=\mathrm{arg}\underset{\mathbf{y}_\mathrm{f}}{\mathrm{min}}\ g\left(\hat{k}_\mathrm{pm}(\eta)\right)\ \mathrm{s.t.}\ m_\mathrm{pm}(\eta)=\mathbf{0},\label{eq:predopt3}
    \end{equation}
    where $g(\cdot)$ can be any optimality measure, such as trace, determinant, or the largest eigenvalue, i.e., the A-, D-, and E-optimality.
\end{remark}
\begin{remark}
    For Hammerstein systems, since $\mathbf{k}_y(\cdot,\cdot)=\mathbf{0}$ as mentioned in Remark~\ref{rm:4}, $k_\mathrm{p}(\eta)$ does not depend on $\mathbf{y}_\mathrm{f}$ and the monotonicity condition $\mathcal{M}$ is automatically satisfied. All optimality criteria admit their minimum when $m_\mathrm{p}(\eta)=\mathbf{0}$, i.e., $m(\eta)-\mathbf{k}^\top(\boldsymbol{\eta}^d,\eta)\left(K+\bar{\Sigma}\right)^{-1}\mathbf{m}(\boldsymbol{\eta}^d)=\mathbf{0}$. This leads to an explicit predictor by using \eqref{eq:m}:
    \begin{equation*}
        \resizebox{\columnwidth}{!}{$\mathbf{y}_\mathrm{f}=\left[\Gamma_1\ \Gamma_2\right]\col{\mathbf{m}_u(\mathbf{u}),\mathbf{y}_\mathrm{p}}-\mathbf{k}^\top(\boldsymbol{\eta}^d,\eta)\left(K+\bar{\Sigma}\right)^{-1}\mathbf{m}(\boldsymbol{\eta}^d).$}
    \end{equation*}
\end{remark}

\subsection{Hyperparameter Tuning}
\label{sec:3c}

There are multiple hyperparameters in the posterior distribution \eqref{eq:distpostm} that require tuning, including kernel parameters $\theta_u$, $\theta_y$, the noise level $\sigma^2$, and linear model parameters $\Gamma_1$, $\Gamma_2$, i.e., $\Theta=\left\{\theta_u,\theta_y,\sigma^2,\Gamma_1,\Gamma_2\right\}$. It is possible to directly solve \eqref{eq:mmlm} for the hyperparameters.

However, the hyperparameters in this problem are high-dimensional, especially since $\left[\Gamma_1\ \Gamma_2\right]\in\mathbb{R}^{n_yL'\times (n_uL+n_yL_0)}$, which is known to induce overfitting \cite{10.7551/mitpress/3206.001.0001}. This problem can be alleviated by selecting a suitable prior distribution for the hyperparameter, known as the hyperprior \cite{Khosravi_2020}. Note that $\Gamma_1$, $\Gamma_2$ correspond to the coefficients of the linear multi-step-ahead predictor for $G(q)$. The problem of designing priors for linear model parameters has been widely studied in the system identification literature. Most prominently, a class of stable spline kernels has been developed; see \cite{pillonetto2022regularized} and references therein.

From \eqref{eq:gam_mdl}, it can be seen that $\Gamma_{12}$ has a block lower-diagonal Toeplitz structure where the last block row corresponds to the system's Markov parameters of length $L'$. For a stable system $G(q)$, a standard stable spline kernel design, such as the TC (tune/correlated) kernel, can be applied as the hyperprior of the last block row. There is no explicit structure for $\Gamma_{11}$ and $\Gamma_2$. In this work, we follow the same idea as in \cite[Section~5.3]{Pillonetto_2014} and model each row of $\Gamma_{11}$ and $\Gamma_2$ with a stable spline kernel indexed by the input-output lag.

For the sake of exposition, we present the case of SISO systems with $n_u=n_y=1$. For MIMO systems, the same hyperprior is enforced on each input-output channel. In detail, Gaussian hyperpriors are considered for $\Gamma_{11}$, $\Gamma_{12}$, and $\Gamma_2$ with zero mean and covariances:
\begin{equation*}
    \resizebox{\columnwidth}{!}{$
    \begin{aligned}
    \cov{\left(\Gamma_{12}\right)_{i,j},\left(\Gamma_{12}\right)_{i',j'}}&=
    \begin{cases}
        0,& j>i\ \mathrm{or}\ j'>i',\\
        s_{i-j,i'-j'},&\mathrm{otherwise}.
    \end{cases}\\
\cov{\left(\Gamma_{11}\right)_{i,j},\left(\Gamma_{11}\right)_{i',j'}}&=
    \begin{cases}
        0,& i\neq i',\\
        s_{L_0+i-j,L_0+i'-j'},&\mathrm{otherwise},
    \end{cases}\\ 
    \cov{\left(\Gamma_2\right)_{i,j},\left(\Gamma_2\right)_{i',j'}}&=
    \begin{cases}
        0,& i\neq i',\\
        s_{L_0+i-j,L_0+i'-j'},&\mathrm{otherwise},
    \end{cases}
    \end{aligned}$}
\end{equation*}
where $s_{i,j}(\zeta)$ is a stable spline kernel parameterized by $\zeta$, which is sometimes known as the hyper-hyperparameters. Throughout this work, the TC kernel is used:
\begin{equation*}
    s^\mathrm{TC}_{i,j}(\zeta) := \lambda\alpha^{\max{(i,j)}},\ \zeta:=[\lambda\ \alpha]^\top,\ \lambda\geq 0,\ 0\leq\alpha<1.
\end{equation*}
The hyper-hyperparameters $\zeta$ can be selected by cross-validation. For alternative stable spline kernel designs, readers are referred to \cite[Section~5.5]{pillonetto2022regularized}.

With hyperpriors introduced for $\Gamma_1$ and $\Gamma_2$, instead of the maximum marginal likelihood problem \eqref{eq:mmlm} which considers all hyperparameters as deterministic variables, a joint maximum-a-posteriori/maximum-likelihood (JMAP-ML) problem is considered \cite{Yeredor_2000}:
\begin{equation}
    \Theta = \mathrm{arg}\underset{\Theta}{\mathrm{max}}\ q\left(\mathcal{M},\boldsymbol{\chi}^d|\boldsymbol{\theta},\Gamma\right)p(\Gamma|\zeta),
    \label{eq:jmapml1}
\end{equation}
where $\boldsymbol{\theta}:=\col{\theta_u,\theta_y,\sigma^2}$ is considered deterministic and $\Gamma:=[\Gamma_1\ \Gamma_2]$ admits a hyperprior.

Vectorizing all the independent elements of $\Gamma$, define $\boldsymbol{\gamma}:=\col{\vect{\Gamma_{11}^\top},\vect{\Gamma_2^\top},\gamma_{12}^\top}$, where $\gamma_{12}$ denotes the last row of the lower-diagonal Toeplitz matrix $\Gamma_{12}$. Then, the hyperprior of $\Gamma$ can be expressed as $\boldsymbol{\gamma}|\zeta\sim\mathcal{N}(\mathbf{0},S_{\boldsymbol{\gamma}})$, where
\begin{multline}
    S_{\boldsymbol{\gamma}}:=\mathrm{blkdiag}(S_{L_0:1},S_{L_0+1:2},\dots,S_{L-1:L'},\\S_{L_0:1},S_{L_0+1:2},\dots,S_{L-1:L'},S_{L'-1:0}),
    \label{eq:sgam}
\end{multline}
where the $(i,j)$-th element of $S_{m:n}\in\mathbb{R}^{(m-n+1)\times (m-n+1)}$ is $s_{m-i+1,m-j+1}$. Therefore, the JMAP-ML problem \eqref{eq:jmapml1} is equivalent to
\begin{multline}
    \bar{\Theta} = \mathrm{arg}\underset{\bar{\Theta}}{\mathrm{min}}\ \logdet{\Lambda_\mathrm{j}(\bar{\Theta})}+\delta_\mathrm{j}^\top \Lambda^{-1}_\mathrm{j}(\bar{\Theta})\delta_\mathrm{j}+\boldsymbol{\gamma}^\top S_{\boldsymbol{\gamma}}^{-1}\boldsymbol{\gamma},
    \label{eq:jmapml2}
\end{multline}
where $\bar{\Theta}:=\left\{\boldsymbol{\theta},\boldsymbol{\gamma}\right\}$ and $\delta_\mathrm{j}$ is defined below \eqref{eq:39}.

\begin{remark}
    Solving \eqref{eq:jmapml2} requires implementing Algorithm~\ref{al:0} inside the optimization problem, which can be time-consuming. For computational efficiency, one can alternatively solve 
    \begin{equation}
        \bar{\Theta} = \mathrm{arg}\underset{\bar{\Theta}}{\mathrm{min}}\ \logdet{\Lambda(\bar{\Theta})}+\delta^\top\Lambda^{-1}(\bar{\Theta})\delta+\boldsymbol{\gamma}^\top S_{\boldsymbol{\gamma}}^{-1}\boldsymbol{\gamma},
        \label{eq:jmapml22}
    \end{equation}
    where $\delta$ is defined below \eqref{eq:mml}, without the virtual derivative points, and only include them in the posterior prediction.
\end{remark}

Algorithm~\ref{al:1} summarizes the procedure of implicit GP data-driven prediction for H-W systems, where lines 2 and 3 estimate the hyperparameters offline and line 5 obtains output predictions online.
\begin{algorithm}[thb]
\caption{Implicit Gaussian process data-driven prediction of Hammerstein-Wiener systems}
\begin{algorithmic}[1]
    \State \textbf{Given:} Hankel data matrices $H_u$, $H_y$, virtual derivative points $\mathbf{y}_\mathrm{m}$, hyper-hyperparameters $\zeta$
    \State Solve \eqref{eq:jmapml2} or \eqref{eq:jmapml22} for $\bar{\Theta}$, where $\delta_\mathrm{j}:=\tilde{\boldsymbol{\mu}}_\mathrm{j}-\mathbf{m}_\mathrm{j}$, $\Lambda_\mathrm{j}(\Theta):=K_\mathrm{j}+\tilde{\Sigma}_\mathrm{j}$ defined in \eqref{eq:monojoint} and \eqref{eq:jstat}, and $S_{\boldsymbol{\gamma}}$ is given by \eqref{eq:sgam}.
    \State Construct $\Gamma_1$, $\Gamma_2$ from $\gamma$.
    \State \textbf{Input:} Prediction condition $\mathbf{u}$, $\mathbf{y}_\mathrm{p}$
    \State Solve \eqref{eq:predopt1}, \eqref{eq:predopt2}, or \eqref{eq:predopt3} for $\mathbf{y}_\mathrm{f}$, where $m_\mathrm{pm}(\eta)$, $k_\mathrm{pm}(\eta)$ are given by \eqref{eq:mpostm} and \eqref{eq:kpostm}, respectively.
    \State \textbf{Output:} Output prediction $\mathbf{y}_\mathrm{f}$
\end{algorithmic}
\label{al:1}
\end{algorithm}

\subsection{Recovering Nonlinearities}
\label{sec:3d}

In addition to output prediction, it could also be of interest to recover the input and output nonlinearities $\psi(\cdot)$ and $\phi(\cdot)$. The output nonlinearity estimate will also be used in Section~\ref{sec:4} for DDPC design.

To obtain $\Phi(\mathbf{y})$ at any vector of query points $\mathbf{y}$, consider the joint distribution:
\begin{equation*}
    \begin{bmatrix}
        \Phi(\mathbf{y})\\\Phi'(\mathbf{y}_\mathrm{m})\\\mathbf{f}(\boldsymbol{\eta}^d)
    \end{bmatrix}\sim\mathcal{N}\left(\begin{bmatrix}
        \mathbf{m}_y(\mathbf{y})\\\mathbf{m}_\mathrm{j}
    \end{bmatrix},\begin{bmatrix}
        \mathbf{k}_y(\mathbf{y},\mathbf{y})&\kappa^\top_{y\mathrm{j}}(\mathbf{y})\\
        \kappa_{y\mathrm{j}}(\mathbf{y})&K_\mathrm{j}
    \end{bmatrix}
    \right),
\end{equation*}
where
\begin{equation*}
    \resizebox{\columnwidth}{!}{$
    \kappa_{y\mathrm{j}}(\mathbf{y})=\col{\frac{\partial \mathbf{k}_y(y_{\mathrm{m},1},\mathbf{y})}{\partial y_{\mathrm{m},1}},\dots,\frac{\partial \mathbf{k}_y(y_{\mathrm{m},n_m},\mathbf{y})}{\partial y_{\mathrm{m},n_m}},\kappa_y(\boldsymbol{\eta}^d,\mathbf{y})}.
    $}
\end{equation*}
Similar to the derivation of \eqref{eq:b6} and \eqref{eq:b7}, this leads to
\begin{equation}
    \Phi(\mathbf{y})|\mathcal{M},\boldsymbol{\chi}^d\sim\mathcal{N}\left(\mathbf{m}_{y\mathrm{p}}(\mathbf{y}),\mathbf{k}_{y\mathrm{p}}(\mathbf{y})\right),
    \label{eq:nlrec1}
\end{equation}
where
\begin{align}
    \mathbf{m}_{y\mathrm{p}}(\mathbf{y})&:=\mathbf{m}_y(\mathbf{y})+\kappa^\top_{y\mathrm{j}}(\mathbf{y})\left(K_\mathrm{j}+\tilde{\Sigma}_\mathrm{j}\right)^{-1}\left(\tilde{\boldsymbol{\mu}}_\mathrm{j}-\mathbf{m}_\mathrm{j}\right),\label{eq:myp}\\
    \mathbf{k}_{y\mathrm{p}}(\mathbf{y})&:=\mathbf{k}_y(\mathbf{y},\mathbf{y})-\kappa^\top_{y\mathrm{j}}(\mathbf{y})\left(K_\mathrm{j}+\tilde{\Sigma}_\mathrm{j}\right)^{-1}\kappa_{y\mathrm{j}}(\mathbf{y}).
\end{align}
Similarly, for $\Psi(\mathbf{u})$, we have
\begin{equation}
    \Psi(\mathbf{u})|\boldsymbol{\chi}^d\sim\mathcal{N}\left(\mathbf{m}_{u\mathrm{p}}(\mathbf{u}),\mathbf{k}_{u\mathrm{p}}(\mathbf{u})\right)
    \label{eq:nlrec2}
\end{equation}
with $\mathbf{m}_{u\mathrm{p}}(\mathbf{u})$ and $\mathbf{k}_{u\mathrm{p}}(\mathbf{u})$ similarly defined and $\kappa_{u\mathrm{j}}(\mathbf{u})=\col{\mathbf{0},\kappa_u(\boldsymbol{\eta}^d,\mathbf{u})}$. Then, $\Psi(\mathbf{u})$ and $\Phi(\mathbf{y})$ are estimated by their posterior means $\mathbf{m}_{u\mathrm{p}}(\mathbf{u})$ and $\mathbf{m}_{y\mathrm{p}}(\mathbf{y})$, respectively.

\section{Data-Driven Predictive Control with Implicit Gaussian Process Model}
\label{sec:4}

This section applies receding horizon DDPC using the implicit GP predictor described in Section~\ref{sec:3}. In detail, considering the problem of tracking an output reference $r_k$ at time $k$, the following expected control cost is considered at time $k$:
\begin{equation}
    J_\mathrm{ctr}:=\norm{\mathbf{u}^k}_R^2 + \mathbb{E}\left[\norm{\Phi(\mathbf{y}^k_0)-\Phi(\mathbf{r}^k)}_Q^2\right],
\end{equation}
where $R\in\mathbb{R}^{n_uL'\times n_uL'}$, $Q\in\mathbb{R}^{n_yL'\times n_yL'}$ are the input and the output cost matrices, respectively, $\mathbf{r}^k:=\col{r_k,\dots,r_{k+L'-1}}$, $\mathbf{u}^k:=\col{u_k,\dots,u_{k+L'-1}}$, $\mathbf{y}^k_0:=\col{y_{k,0},\dots,y_{k+L'-1,0}}$ are the reference, input, and true noise-free output trajectories at time $k$, respectively, and $L'$ is the prediction horizon. In addition, we also consider input constraints $\mathbf{u}^k\in\mathcal{U}_k$ and linear output constraints $H\mathbf{y}^k_0\leq q$, where $H\in\mathbb{R}^{n_c\times n_yL'}$ and $q\in\mathbb{R}^{n_c}$. Since the prediction error is unbounded, the output constraint can only be enforced with high probability $p$ as a chance constraint $\pr{H\mathbf{y}^k_0\leq q}\geq p$.

In this section, the following continuity assumption on the output nonlinearity is considered.
\begin{assumption}
     The output nonlinearity $\phi^{-1}(\cdot)$ is Lipschitz continuous with $\norm{\phi^{-1}(\bar{y}_1)-\phi^{-1}(\bar{y}_2)}_2\leq M\norm{\bar{y}_1-\bar{y}_2}_2$, where $M$ is the Lipschitz constant.
     \label{ass:2}
\end{assumption}

As in standard predictive control, an explicit output prediction of $\mathbf{y}^k_0$ could be obtained by applying Algorithm~\ref{al:1} to a particular prediction condition $\mathbf{u}=\col{u_{k-L_0},\dots,u_{k-1},\mathbf{u}^k},\ \mathbf{y}_\mathrm{p}=\col{y_{k-L_0},\dots,y_{k-1}}$. However, this requires solving \eqref{eq:predopt1}, \eqref{eq:predopt2}, or \eqref{eq:predopt3} as an inner problem, which is computationally challenging. Instead, we directly utilize the implicit predictor by quantifying the prediction error for an arbitrary input-output trajectory
\begin{equation*}
    \eta^k:=\col{u_{k-L_0},\dots,u_{k-1},\mathbf{u}^k,y_{k-L_0},\dots,y_{k-1},\mathbf{y}^k}.
\end{equation*}
From \eqref{eq:pperror}, the distribution of $\Phi(\mathbf{y}^k_0)$ is given by
\begin{equation}
    \Phi(\mathbf{y}^k_0)\sim\mathcal{N}\left(\Phi(\mathbf{y}^k)+m_\mathrm{pm}(\eta^k),\hat{k}_\mathrm{pm}(\eta^k)\right),
    \label{eq:phidist}
\end{equation}
so the expected control cost can be calculated as
\begin{multline*}
    J_\mathrm{ctr}=\norm{\mathbf{u}^k}_R^2 + \norm{\Phi(\mathbf{y}^k)+m_\mathrm{pm}(\eta^k)-\Phi(\mathbf{r}^k)}_Q^2\\+\tr{Q\hat{k}_\mathrm{pm}(\eta^k)}.
\end{multline*}
Since the output nonlinearity $\Phi(\cdot)$ is not known exactly, $\Phi(\mathbf{y}^k)$ and $\Phi(\mathbf{r}^k)$ are replaced by their posterior means $\mathbf{m}_{y\mathrm{p}}(\mathbf{y}^k)$ and $\mathbf{m}_{y\mathrm{p}}(\mathbf{r}^k)$ with certainty equivalence.

Then, the control problem optimizes $\mathbf{u}^k$ and $\mathbf{y}^k$ at the same time without considering the dependence of $\mathbf{u}^k$ and $\mathbf{y}^k$ explicitly.

\begin{remark}
    In standard GP-MPC, a one-step-ahead GP predictor is applied recursively, whereas a multi-step-ahead predictor is directly applied here. Multi-step-ahead predictors generally induce less conservatism in MPC and lead to less complex problems \cite{Kohler_2022}. In particular, for GP-MPC, our approach circumvents the difficult problem of uncertainty propagation since the posterior of a GP from an input distribution is generally intractable, and the resulting distribution is not Gaussian \cite{Hewing_2020}.
\end{remark}
Finally, the chance constraint satisfaction is guaranteed by the following lemma.
\begin{lemma}
    Under Assumptions~\ref{ass:1} and \ref{ass:2}, 
    the chance constraint $\pr{H\mathbf{y}^k_0\leq q}\geq p$ is satisfied if
    \begin{equation}
        H\mathbf{y}^k\leq q - c_p\sqrt{\diag{HH^\top}},
        \label{eq:confinal}
    \end{equation}
    where $c_p:=M\left(\sqrt{\mu_{\chi^2}(p)\sigma_\mathrm{p}(\eta^k)}+\norm{m_\mathrm{pm}(\eta^k)}_2\right)$, $\sigma_\mathrm{p}(\eta^k)$ is the largest eigenvalue of $\hat{k}_\mathrm{pm}(\eta^k)$ and $\mu_{\chi^2}(\cdot)$ is the quantile function of the $\chi^2$-distribution with $n_yL'$ degrees of freedom.
\end{lemma}
\begin{pf}
From \eqref{eq:phidist}, we have
\begin{equation*}
    \pr{\norm{\Phi(\mathbf{y}^k_0)-\Phi(\mathbf{y}^k)-m_\mathrm{pm}(\eta^k)}^2_{\hat{k}_\mathrm{pm}^{-1}(\eta^k)}\leq\mu_{\chi^2}(p)}= p,
\end{equation*}
since $\norm{\Phi(\mathbf{y}^k_0)-\Phi(\mathbf{y}^k)-m_\mathrm{pm}(\eta^k)}^2_{\hat{k}_\mathrm{pm}^{-1}(\eta^k)}$ follows a $\chi^2$-distribution with $n_yL'$ degrees of freedom. Using Assumption~\ref{ass:2} and the reverse triangle inequality, we have
\begin{equation*}
    \begin{split}
    &\norm{\Phi(\mathbf{y}^k_0)-\Phi(\mathbf{y}^k)-m_\mathrm{pm}(\eta^k)}_{\hat{k}_\mathrm{pm}^{-1}(\eta^k)}\\
    \geq&\ \tfrac{1}{\sqrt{\sigma_\mathrm{p}(\eta^k)}}\norm{\Phi(\mathbf{y}^k_0)-\Phi(\mathbf{y}^k)-m_\mathrm{pm}(\eta^k)}_2\\
    \geq&\ \tfrac{1}{\sqrt{\sigma_\mathrm{p}(\eta^k)}}\left(\norm{\Phi(\mathbf{y}^k_0)-\Phi(\mathbf{y}^k)}_2-\norm{m_\mathrm{pm}(\eta^k)}_2\right)\\
    \geq&\ \tfrac{1}{M\sqrt{\sigma_\mathrm{p}(\eta^k)}}\norm{\mathbf{y}^k_0-\mathbf{y}^k}_2-\tfrac{1}{\sqrt{\sigma_\mathrm{p}(\eta^k)}}\norm{m_\mathrm{pm}(\eta^k)}_2.
    \end{split}
\end{equation*}
This leads to $\pr{\norm{\mathbf{y}^k_0-\mathbf{y}^k}_2\leq c_p}\geq p$. If $\norm{\mathbf{y}^k_0-\mathbf{y}^k}_2\leq c_p$, according to \cite[Theorem 2.3]{kolmanovsky1998theory}, $H\mathbf{y}^k_0\leq q$ is guaranteed by
\begin{equation}
    h_i^\top\mathbf{y}^k\leq q_i-\rho_V(h_i),\quad i=1,\dots,n_c,
    \label{eq:hitight}
\end{equation}
where $H=:\col{h_1^\top,\dots,h_{n_c}^\top}$, $q=:\col{q_1,\dots,q_{n_c}}$, and $\rho_V(h_i)=c_p\sqrt{h_i^\top h_i}$ denotes the support function of the set $V:=\left\{\mathbf{e}\,|\norm{\mathbf{e}}_2\leq c_p\right\}$. Aggregating \eqref{eq:hitight} for all $i$ directly leads to the lemma.\qed
\end{pf}

To sum up, the following optimization problem is solved at time $k$:
\begin{equation}
    \begin{aligned}
        \underset{\mathbf{u}^k,\mathbf{y}^k}{\mathrm{min}}&\norm{\mathbf{u}^k}_R^2 + \norm{\mathbf{m}_{y\mathrm{p}}(\mathbf{y}^k)+m_\mathrm{pm}(\eta^k)-\mathbf{m}_{y\mathrm{p}}(\mathbf{r}^k)}_Q^2\\
        &\qquad\qquad\qquad\qquad\qquad\qquad\quad\ \ +\tr{Q\hat{k}_\mathrm{pm}(\eta^k)}\\
        \mathrm{s.t.}\ &\ \mathbf{u}^k\in\mathcal{U}_k,\ H\mathbf{y}^k\leq q - c_p\sqrt{\diag{HH^\top}},
    \end{aligned}
\label{eq:ctr}
\end{equation}
where $\mathbf{m}_{y\mathrm{p}}(\cdot)$ is the estimated output nonlinearity given by \eqref{eq:myp}, $m_\mathrm{pm}(\cdot)$ is the posterior mean of the implicit function given by \eqref{eq:mpostm}, and $\hat{k}_\mathrm{pm}(\cdot)$ is the modified posterior covariance of the implicit function given by \eqref{eq:kpostm} and \eqref{eq:khat}. Then, the first entry in the optimal input sequence is applied to the system, and the noisy output $y_k$ is measured.

As discussed above, there is no explicit predictor in \eqref{eq:ctr}. Instead, the implicit predictor is embedded in the expected cost function and the constraint tightening term, penalizing both the bias ($m_\mathrm{pm}(\eta^k)$) and the variance ($\hat{k}_\mathrm{pm}(\eta^k)$) errors. This implicitly prohibits unreliable choices of $\eta^k$, which are unlikely to be generated by the system.

\section{Numerical Examples}
\label{sec:5}

This section demonstrates the performance of the proposed data-driven predictor (Algorithm~\ref{al:1}) and controller \eqref{eq:ctr}.\footnote{The codes are available at https://doi.org/10.25835/ 0ahxlrtw.}

The following parameters are used throughout this section: $L_0=n_x$, $N=100$, $\sigma=0.01$. The hyper-hyperparameters $\zeta=[\lambda\ \alpha]^\top$ are selected by cross-validation with grid $\lambda\in\{0.25,0.5,1,2,4\}$, $\alpha\in\{0.5,0.6,0.7,0.8,0.9\}$. The squared exponential kernel is used for $k_u(\cdot,\cdot)$ and $k_y(\cdot,\cdot)$, i.e., $k_u(u',u'';l_u)=\exp\left(-\frac{(u'-u'')^2}{2l_u^2}\right)$ with hyperparameter $l_u$, and similarly for $k_y(y',y'';l_y)$. Note that the scaling factor is omitted as discussed in Remark~\ref{rm:40}. The prior mean functions are selected as identity, i.e., $m_u(u)=u$, $m_y(y)=y$. The minimum mean-squared error objective \eqref{eq:predopt2} is used to obtain explicit predictions in Algorithm~\ref{al:1} and solved using the \textsc{Matlab} function \texttt{particleswarm} with a particle swarm solver. The other optimization problems \eqref{eq:jmapml2} and \eqref{eq:ctr} are solved using the \textsc{Matlab} function \texttt{fmincon} with the interior point algorithm.

The proposed implicit GP model is benchmarked against a black-box GP model without incorporating the H-W model structure. To obtain multi-step-ahead predictions, denoted by
$$y_{\mathrm{f},l}\sim\mathcal{N}\left(m_{\mathrm{bb},l}(\col{\mathbf{u},\mathbf{y}_\mathrm{p}}),\sigma_{\mathrm{bb},l}^2(\col{\mathbf{u},\mathbf{y}_\mathrm{p}})\right),$$
where $m_{\mathrm{bb},l}(\cdot)$, $\sigma_{\mathrm{bb},l}(\cdot)$ are the posterior nominal prediction and standard deviation of the $l$-step-ahead predictor, respectively, two strategies are considered. 1) Black-box posterior predictors for all prediction horizons $l=1,\dots,L'$ are learned separately. 2) The one-step-ahead black-box predictor is applied recursively with nominal predictions propagated with certainty equivalence. The models are trained using the \textsc{Matlab} function \texttt{fitrgp} with hyperparameters estimated by \texttt{bayesopt} using 100 evaluations.

We also compare a linear predictor based on WFL, neglecting the nonlinearities: 
$$\mathbf{y}_\mathrm{f}=H_{y\mathrm{f}}\,\col{H_u,H_{y\mathrm{p}}}^\dagger\col{\mathbf{u},\mathbf{y}_\mathrm{p}},$$
where $H_y=:\col{H_{y\mathrm{p}},H_{y\mathrm{f}}}$, $H_{y\mathrm{p}}\in\mathbb{R}^{n_y L_0\times M}$, and $H_{y\mathrm{f}}\in\mathbb{R}^{n_y L'\times M}$.

\subsection{Data-Driven Prediction}

We first benchmark the performance of Algorithm~\ref{al:1}. In this subsection, the linear part of the model $G(q)$ is selected as a random single-input, single-output second-order system, generated by the \textsc{Matlab} function \texttt{drss}. The systems are normalized to have an $\mathcal{H}_2$-norm of 10. Gaussian inputs with a standard deviation of 2 are used for both training and testing.

For demonstrative purposes, we first investigate one-step-ahead prediction with $L'=1$ and fixed input and output nonlinearities $\psi(u)=u+\sin(u)$, $\phi^{-1}(\bar{y})=\bar{y}+\sin(\bar{y})$. We consider two cases: 1) with 20 virtual derivative points in a linear grid between $-\pi$ and $\pi$ and 2) without virtual derivative points $n_m=0$, indicated by ``w/o mono.'' in the figures.

\begin{figure}
    \centering
    \includegraphics[width=\columnwidth]{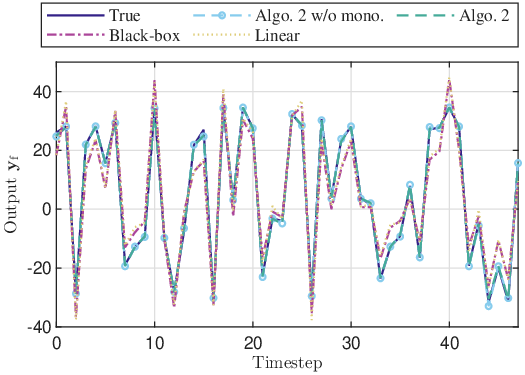}
    \caption{Comparison of data-driven prediction with different predictors. W/o mono.: no virtual derivative points.}
    \label{fig:1}
\end{figure}

\begin{figure}
    \centering
    \includegraphics[width=\columnwidth]{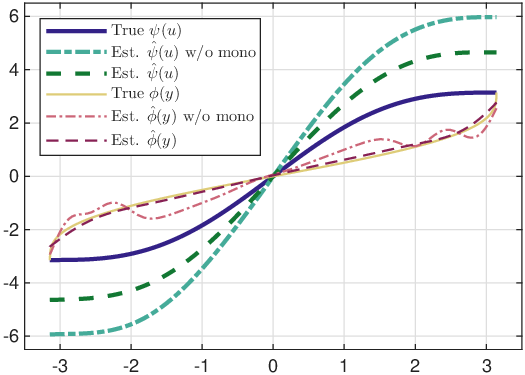}
    \caption{Estimation of input and output nonlinear functions $\psi(u)$ and $\phi(y)$. W/o mono.: no virtual derivative points.}
    \label{fig:2}
\end{figure}

Figure~\ref{fig:1} demonstrates the prediction accuracy of the predictors on a 50-point test dataset. The recovered input and output nonlinear functions are plotted in Figure~\ref{fig:2} by applying \eqref{eq:nlrec1}-\eqref{eq:nlrec2} to a linear grid of query points between $-\pi$ and $\pi$. It can be seen from Figure~\ref{fig:1} that Algorithm~\ref{al:1} obtains very accurate predictions and performs much better than the black-box GP model and the linear predictor. The addition of virtual derivative points does not have a large effect on the predictions. The black-box predictions are similar to the linear ones, indicating that the black-box GP model does not capture the nonlinearities well. As can be seen from Figure~\ref{fig:2}, the shapes of the nonlinearities are captured accurately by Algorithm~\ref{al:1} with virtual derivative points, whereas the estimated output nonlinearity is not monotonic when virtual derivative points are absent. This proves the effectiveness of promoting monotonicity by introducing virtual derivative points. There is a discrepancy of $\psi(u)$ in terms of the scaling, which is expected since only the normalized nonlinearities are recovered, cf. Remark~\ref{rm:40}.


Then, multi-step-ahead predictors with $L'=4$ are considered by conducting 100 Monte Carlo simulations. The configurations are the same as the one-step-ahead case, except that random input and output nonlinearities are considered by generating samples from the following Gaussian process: $f(\cdot)\sim\mathcal{GP}\left(m(\cdot),k(\cdot,\cdot)\right)$, $m(\eta)=\eta$, $k(\eta',\eta'')=\exp\left(-(\eta'-\eta'')^2/8\right)$, and the virtual derivative point are selected between $-$5 and 5. The output nonlinearity is verified to be monotonically increasing.

\begin{figure}
    \centering
    \includegraphics[width=\columnwidth]{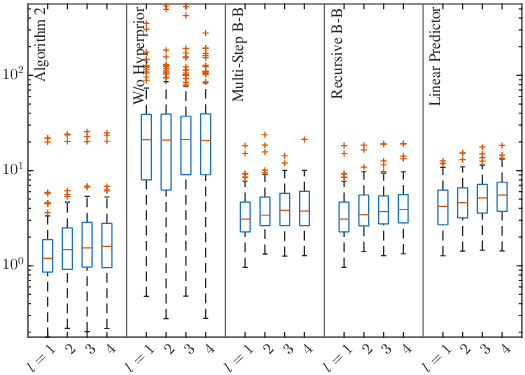}
    \caption{Boxplot of root-mean-squared prediction errors for different predictors with $L'=4$. $l$: step of forward prediction.}
    \label{fig:33}
\end{figure}

At each time $k$, $l$-step-ahead prediction accuracy is evaluated quantitatively on a 50-point test data set by comparing the $l$-step-ahead prediction at time $k$ with the true output at time $(k+l)$, $l=1,2,3,4$. The boxplot of root-mean-squared prediction errors is shown in Figure~\ref{fig:33}. In addition to the predictors presented before, we also consider the implicit GP model with hyperparameters estimated without the hyperprior, i.e., $S_\gamma^{-1}=\mathbf{0}$ in \eqref{eq:jmapml2}. This is indicated by ``w/o hyperprior'' in Figure~\ref{fig:33}.

Similar to one-step-ahead prediction, our proposed algorithm achieves the smallest prediction errors over the whole horizon, with overall median error reductions of 59.8\%, 58.8\%, and 70.5\% compared to the multi-step black-box GP model, the recursive one-step GP model, and the linear predictor, respectively. On the other hand, a significant prediction error is observed when the hyperprior is not used in hyperparameter estimation. This indicates that the proposed method would fail without the hyperprior, validating the effectiveness of the proposed JMAP-ML scheme for hyperparameter estimation.

Admittedly, our proposed approach suffers from high computational complexity, with a median training time of 69\,s for each hyper-hyperparameter choice and a median prediction time of 56\,s for 50 predictions on an AMD Ryzen 7 8845H laptop. In comparison, the black-box GP model requires a median training time of 46\,s and a median prediction time of 0.02\,s.

\subsection{Data-Driven Predictive Control}

Finally, the DDPC algorithm \eqref{eq:ctr} is applied with prediction horizon $L'=4$. In this subsection, we consider the following pH process model obtained in \cite{Abinayadhevi_2015}:
\begin{align*}
    G(q)&=\frac{-1.728q^3+q^2+0.5056q}{q^3-0.797q^2-0.1609q+0.06203},\\
    \psi(u)&=\frac{1-e^{-u}}{1+e^{-u}},\quad\phi^{-1}(\bar{y})=\frac{1-e^{-\bar{y}}}{1+e^{-\bar{y}}}.
\end{align*}
The output nonlinearity is monotonically increasing and satisfies Assumption~\ref{ass:2} with $M=0.5$. Unit Gaussian inputs are used to obtain the training data. A sinusoidal output reference with lower and upper bound constraints $-0.2\leq\mathbf{y}^k-\mathbf{r}^k\leq 0.2$ is considered. No input constraints are enforced, i.e., $\mathcal{U}_k=\mathbb{R}^{L'}$. The following parameters are used: $M=0.5$, $p=0.7$, $Q=100\cdot\mathbb{I}$, $R=\mathbb{I}$.

We have also implemented receding horizon control for the black-box GP and the linear predictors. For the black-box GP models, the following optimization problem is solved at time $k$:
\begin{equation*}
\begin{aligned}
    \underset{\mathbf{u}^k,\mathbf{y}^k}{\mathrm{min}}&\ \norm{\mathbf{u}^k}_R^2 + \norm{\mathbf{y}^k-\mathbf{r}^k}_Q^2+\diag{Q}^\top\boldsymbol{\sigma}^2_\mathrm{bb}(\xi^k)\\
    \mathrm{s.t.}\ &\ \mathbf{y}^k=\mathbf{m}_\mathrm{bb}(\xi^k),\ \mathbf{u}^k\in\mathcal{U}_k,\ \\
    &\ H\mathbf{y}^k\leq q - \sqrt{\mu_{\chi^2}(p)\diag{H\Sigma_\mathrm{bb}(\xi^k)H^\top}},
\end{aligned}
\end{equation*}
where $\xi^k=\col{u_{k-L_0},\dots,u_{k-1},\mathbf{u}^k,y_{k-L_0},\dots,y_{k-1}}$ and $\Sigma_\mathrm{bb}(\xi^k)$ denotes the diagonal matrix constructed from $\boldsymbol{\sigma}_\mathrm{bb}^2(\xi^k)$. The derivation is similar to Section~\ref{sec:4}. To quantify the multi-step prediction error for the recursive GP model, we tried to apply the linear approximation approach \cite{Hewing_2020}, but this leads to infeasible problems at all times due to too conservative error quantification. Instead, we ignore the prediction error due to noisy regression inputs. For the linear predictor, we apply subspace predictive control \cite{Fiedler_2021} with certainty equivalence. In addition, we have applied nonlinear MPC using the true nonlinear model as a benchmark for ideal performance.

\begin{figure*}
    \centering
    \includegraphics[width=\textwidth]{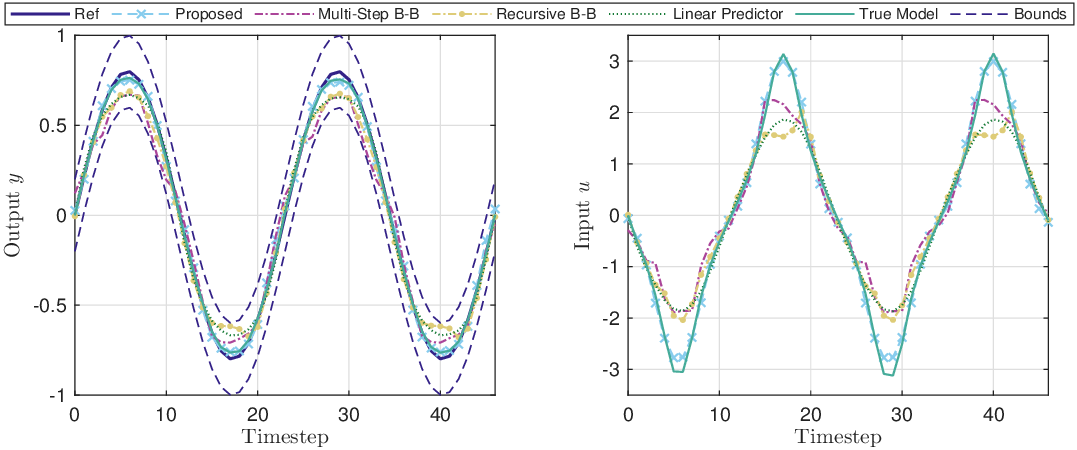}
    \caption{Closed-loop input-output trajectories with different data-driven predictive controllers.}
    \label{fig:4}
\end{figure*}

The closed-loop input-output trajectories are plotted in Figure~\ref{fig:4}. The results demonstrate that the proposed control algorithm \eqref{eq:ctr} achieves the best performance among the data-driven algorithms. It performs very similarly to the nonlinear MPC algorithm with the true model and is able to track the reference closely within the output bounds. The black-box GP algorithms and the linear predictor under-actuate at the peaks of the output reference, indicating that they are not accurately capturing the output nonlinearity of the model.

Similar to prediction, our proposed method requires a longer computation time. The median computation time per iteration is 4.41\,s, 0.24\,s, 0.13\,s, 0.30\,s, and 0.01\,s for the DDPC algorithm \eqref{eq:ctr}, multi-step black-box GP, recursive black-box GP, subspace predictive control, and the nonlinear MPC, respectively.

\section{Conclusions}
\label{sec:6}

This work presents data-driven prediction and control algorithms of Hammerstein-Wiener systems by implicit GP regression. By formulating the predictor as an implicit function, a Gaussian process (GP) model is obtained by deriving kernel functions with linear model parameters and GP priors for nonlinearities. A stable spline hyperprior is used to estimate linear model parameters as hyperparameters. Monotonicity information of the nonlinearities can be included by expectation propagation with virtual derivative points. Applied to receding horizon control, the algorithm minimizes the expected control cost and complies with output chance constraints. Validation through numerical examples demonstrates improved performance over black-box GP methods.

One drawback of the proposed approach is its high computational complexity due to a high number of hyperparameters and complex kernel design. Future work may discuss tailored algorithms for solving the optimization problems. Closed-loop guarantees for the receding horizon control algorithm can also be investigated.

\bibliographystyle{plain}
\bibliography{refs}

\appendix
\section{Derivation of \eqref{eq:mfinal}-\eqref{eq:Kfinal}}
\label{app:1}

Let
$$\mathbf{u}_k^d=\col{u^d_k,\dots,u^d_{k+L-1}},\ \mathbf{y}_k^d=\col{y^d_k,\dots,y^d_{k+L-1}},$$
such that $\vect{H_u}=\col{\mathbf{u}_1^d,\dots,\mathbf{u}_M^d}$ and $\vect{H_y}=\col{\mathbf{y}_1^d,\dots,\mathbf{y}_M^d}$. From \eqref{eq:tdata} and \eqref{eq:m}, we have
\begin{equation*}
    m(\eta_k^d)=\left[\Gamma_1\ \bar{\Gamma}_2\right]\col{\mathbf{m}_u(\mathbf{u}_k^d),\mathbf{m}_y(\mathbf{y}_k^d)}.
\end{equation*}
This leads to
\begin{equation*}
    \left[m(\eta^d_1)\ \cdots\ m(\eta^d_M)\right]=\left[\Gamma_1\ \bar{\Gamma}_2\right]\begin{bmatrix}
    \mathbf{m}_u(H_u)\\\mathbf{m}_y(H_y)
\end{bmatrix}.
\end{equation*}
Vectorizing both sides leads to \eqref{eq:mfinal}.

From \eqref{eq:tdata} and \eqref{eq:priork}, we have
\begin{equation*}
    k(\eta_k^d,\eta)=\Gamma_1\mathbf{k}_u(\mathbf{u}_k^d,\mathbf{u})\Gamma_1^\top+\bar{\Gamma}_2\mathbf{k}_y(\mathbf{y}_k^d,\mathbf{y})\bar{\Gamma}_2^\top.
\end{equation*}
This leads to
\begin{equation*}
\resizebox{\columnwidth}{!}{$
    \begin{aligned}
    &\mathbf{k}(\boldsymbol{\eta}^d,\eta)=\col{k(\eta^d_1,\eta),\dots,k(\eta^d_M,\eta)}\\
    =&\left(\mathbb{I}\otimes\Gamma_1\right)\begin{bmatrix}
        \mathbf{k}_u(\mathbf{u}_1^d,\mathbf{u})\\
        \vdots\\
        \mathbf{k}_u(\mathbf{u}_M^d,\mathbf{u})
    \end{bmatrix}\Gamma_1^\top+\left(\mathbb{I}\otimes\bar{\Gamma}_2\right)\begin{bmatrix}
        \mathbf{k}_y(\mathbf{y}_1^d,\mathbf{y})\\
        \vdots\\
        \mathbf{k}_y(\mathbf{y}_M^d,\mathbf{y})
    \end{bmatrix}\bar{\Gamma}_2^\top\\
    =&\left(\mathbb{I}\otimes\Gamma_1\right)\mathbf{k}_u(\vect{H_u},\mathbf{u})\Gamma_1^\top+\left(\mathbb{I}\otimes \bar{\Gamma}_2\right)\mathbf{k}_y(\vect{H_y},\mathbf{y})\bar{\Gamma}_2^\top.
    \end{aligned}$}
\end{equation*}
Similarly,
\begin{equation*}
    k(\eta_k^d,\eta_k^d)=\Gamma_1\mathbf{k}_u(\mathbf{u}_k^d,\mathbf{u}_k^d)\Gamma_1^\top+\bar{\Gamma}_2\mathbf{k}_y(\mathbf{y}_k^d,\mathbf{y}_k^d)\bar{\Gamma}_2^\top,
\end{equation*}
and
\begin{equation*}
\begin{aligned}
    &K(\boldsymbol{\eta}^d,\boldsymbol{\eta}^d)\\
    =&\left(\mathbb{I}\otimes\Gamma_1\right)\begin{bmatrix}
        \mathbf{k}_u(\mathbf{u}_1^d,\mathbf{u}_1^d)&\cdots&\mathbf{k}_u(\mathbf{u}_1^d,\mathbf{u}_M^d)\\
        \vdots&\ddots&\vdots\\
            \mathbf{k}_u(\mathbf{u}_M^d,\mathbf{u}_1^d)&\cdots&\mathbf{k}_u(\mathbf{u}_M^d,\mathbf{u}_M^d)
    \end{bmatrix}\left(\mathbb{I}\otimes\Gamma_1^\top\right)\\
    +&\left(\mathbb{I}\otimes\bar{\Gamma}_2\right)\begin{bmatrix}
        \mathbf{k}_y(\mathbf{y}_1^d,\mathbf{y}_1^d)&\cdots&\mathbf{k}_y(\mathbf{y}_1^d,\mathbf{y}_M^d)\\
        \vdots&\ddots&\vdots\\
            \mathbf{k}_y(\mathbf{y}_M^d,\mathbf{y}_1^d)&\cdots&\mathbf{k}_y(\mathbf{y}_M^d,\mathbf{y}_M^d)
    \end{bmatrix}\left(\mathbb{I}\otimes\bar{\Gamma}_2^\top\right)\\
    =&\left(\mathbb{I}\otimes\Gamma_1\right)\mathbf{k}_u(\vect{H_u},\vect{H_u})\left(\mathbb{I}\otimes\Gamma_1^\top\right)\\
    &\qquad+\left(\mathbb{I}\otimes \bar{\Gamma}_2\right)\mathbf{k}_y(\vect{H_y},\vect{H_y})\left(\mathbb{I}\otimes \bar{\Gamma}_2^\top\right).
\end{aligned}
\end{equation*}

\section{Details of the expectation propagation algorithm}
\label{app:2}
Let $F_\mathcal{N}\left(\phi'(y_{\mathrm{m},i})/\nu_0\right)$ be approximated by $t_i(\tilde{Z}_i,\tilde{\mu}_i,\tilde{\sigma}_i^2):=\tilde{Z}_i p_{\mathcal{N}(\tilde{\mu}_i,\tilde{\sigma}_i^2)}(\phi'(y_{\mathrm{m},i}))$,
where $p_{\mathcal{N}(\tilde{\mu}_i,\tilde{\sigma}_i^2)}(\phi'(y_{\mathrm{m},i}))$ denotes the probability density function of $\mathcal{N}(\tilde{\mu}_i,\tilde{\sigma}_i^2)$. Let
\begin{equation}
    \begin{aligned}
    \tilde{\boldsymbol{\mu}}_\mathrm{j}&:=\col{\tilde{\mu}_1,\dots,\tilde{\mu}_{n_m},\boldsymbol{\chi}^d=\mathbf{0}},\\
    \tilde{\Sigma}_\mathrm{j}&:=\blkdiag{\tilde{\sigma}_1^2,\dots,\tilde{\sigma}_{n_m}^2,\bar{\Sigma}},\\
    \mathbf{z}&:=\col{\Phi'(\mathbf{y}_\mathrm{m}),\mathbf{f}(\boldsymbol{\eta}^d)}.
    \end{aligned}
    \label{eq:jstat}
\end{equation}
Define the approximate posterior distribution:
\begin{equation}
    q(\mathbf{z}|\mathcal{M},\boldsymbol{\chi}^d):=\frac{1}{Z} p(\mathbf{z})p(\boldsymbol{\chi}^d|\mathbf{f}(\boldsymbol{\eta}^d))\prod_{i=1}^{n_m}t_i(\tilde{Z}_i,\tilde{\mu}_i,\tilde{\sigma}_i^2),
\end{equation}
where $Z$ is the normalization term and $q(\cdot)$ denotes approximate probability densities. We would like to find parameters $\tilde{Z}_i$, $\tilde{\mu}_i$, and $\tilde{\sigma}_i^2$ such that
$$q(\mathbf{z}|\mathcal{M},\boldsymbol{\chi}^d)\approx p(\mathbf{z}|\mathcal{M},\boldsymbol{\chi}^d).$$
Note that
$$p(\boldsymbol{\chi}^d|\mathbf{f}(\boldsymbol{\eta}^d))\prod_{i=1}^{n_m}t_i(\tilde{Z}_i,\tilde{\mu}_i,\tilde{\sigma}_i^2)=p_{\mathcal{N}(\tilde{\boldsymbol{\mu}}_\mathrm{j},\tilde{\Sigma}_\mathrm{j})}(\mathbf{z}).$$
Then, from \eqref{eq:eppost} and the Gaussian product rule, we have approximately
\begin{equation}
    \mathbf{z}|\mathcal{M},\boldsymbol{\chi}^d\sim \mathcal{N}\left(\boldsymbol{\mu},\mathbf{S}\right),
    \label{eq:postm1}
\end{equation}
where
\begin{equation*}
    \boldsymbol{\mu}=\mathbf{S}\tilde{\Sigma}_\mathrm{j}^{-1}\tilde{\boldsymbol{\mu}}_\mathrm{j}+\mathbf{S}K_\mathrm{j}^{-1}\mathbf{m}_\mathrm{j},\quad \mathbf{S}=(K_\mathrm{j}^{-1}+\tilde{\Sigma}_\mathrm{j}^{-1})^{-1}.
\end{equation*}
To achieve this, the EP algorithm starts from some initial approximate posterior and calculates the so-called \textit{cavity distribution} for a particular virtual derivative point $y_{m,i}$. The cavity distribution is defined as the marginal posterior distribution $q(\phi'(y_{\mathrm{m},i})|\mathcal{M},\boldsymbol{\chi}^d)$ divided by the approximant $t_i(\tilde{Z}_i,\tilde{\mu}_i,\tilde{\sigma}_i^2)$. Then we approximate the product of the cavity distribution and the likelihood $F_\mathcal{N}\left(\phi'(y_{\mathrm{m},i})/\nu_0\right)$ with a Gaussian likelihood by matching their zero-th, first, and second moments. Finally, the approximant is updated such that the product of the cavity distribution and the approximant has the desired moments. This process iterates over all the virtual derivative points until convergence.

With the augmentation of the monotonicity information at the virtual derivative points, for any query point $\eta=\col{\mathbf{u},\mathbf{y}}$, we have approximately
\begin{equation}
    \begin{bmatrix}
        f(\eta)\\\mathbf{z}
    \end{bmatrix}\sim\mathcal{N}\left(\begin{bmatrix}
        m(\eta)\\\mathbf{m}_\mathrm{j}
    \end{bmatrix},\begin{bmatrix}
        k(\eta,\eta)&\mathbf{k}^\top_\mathrm{j}(\eta)\\
        \mathbf{k}_\mathrm{j}(\eta)&K_\mathrm{j}
    \end{bmatrix}
    \right),
    \label{eq:jointm}
\end{equation}
where $\mathbf{k}_\mathrm{j}(\eta):=\col{{\kappa'_y}^\top(\eta,\mathbf{y}_\mathrm{m}),\mathbf{k}(\boldsymbol{\eta}^d,\eta)}$.
This leads to the conditional probability:
\begin{multline}
    f(\eta)|\mathbf{z}\sim\mathcal{N}\Big(m(\eta)+\mathbf{k}^\top_\mathrm{j}(\eta)K_\mathrm{j}^{-1}(\mathbf{z}-\mathbf{m}_\mathrm{j}),\\k(\eta,\eta)-\mathbf{k}^\top_\mathrm{j}(\eta)K_\mathrm{j}^{-1}\mathbf{k}_\mathrm{j}(\eta)\Big).
    \label{eq:b5}
\end{multline}
From the definition of conditional probability, we have
\begin{equation*}
q\left(f(\eta)|\mathcal{M},\boldsymbol{\chi}^d\right)=p\left(f(\eta)|\mathbf{z}\right)q(\mathbf{z}|\mathcal{M},\boldsymbol{\chi}^d).
\end{equation*}
From \eqref{eq:postm1} and \eqref{eq:b5}, the posterior mean and covariance of $f(\eta)|\mathcal{M},\boldsymbol{\chi}^d$ are given by
\begin{equation}
    \begin{aligned}
    &\mathbb{E}\left[f(\eta)|\mathcal{M},\boldsymbol{\chi}^d\right]=\int\mathbb{E}\left[f(\eta)|\mathbf{z}\right]q(\mathbf{z}|\mathcal{M},\boldsymbol{\chi}^d)\mathrm{d}\mathbf{z}\\
    =\ &m(\eta)-\mathbf{k}^\top_\mathrm{j}(\eta)K_\mathrm{j}^{-1}\mathbf{m}_\mathrm{j}+\mathbf{k}^\top_\mathrm{j}(\eta)K_\mathrm{j}^{-1}\boldsymbol{\mu}\\
    =\ &m(\eta)+\mathbf{k}^\top_\mathrm{j}(\eta)(K_\mathrm{j}+\tilde{\Sigma}_\mathrm{j})^{-1}\left(\tilde{\boldsymbol{\mu}}_\mathrm{j}-\mathbf{m}_\mathrm{j}\right),
    \end{aligned}
    \label{eq:b6}
\end{equation}
\begin{equation}
    \begin{aligned}
    &\cov{f(\eta)|\mathcal{M},\boldsymbol{\chi}^d}\\
    =\ &k(\eta,\eta)-\mathbf{k}^\top_\mathrm{j}(\eta)K_\mathrm{j}^{-1}\mathbf{k}_\mathrm{j}(\eta)+\mathbf{k}^\top_\mathrm{j}(\eta)K_\mathrm{j}^{-1}SK_\mathrm{j}^{-1}\mathbf{k}_\mathrm{j}(\eta)\\
    =\ &k(\eta,\eta)-\mathbf{k}^\top_\mathrm{j}(\eta)(K_\mathrm{j}+\tilde{\Sigma}_\mathrm{j})^{-1}\mathbf{k}_\mathrm{j}(\eta),
    \end{aligned}
    \label{eq:b7}
\end{equation}
where we used the property:
$$\left(K_\mathrm{j}+\tilde{\Sigma}_\mathrm{j}\right)^{-1}=K_\mathrm{j}^{-1}\left(K_\mathrm{j}^{-1}+\tilde{\Sigma}_\mathrm{j}^{-1}\right)^{-1}\tilde{\Sigma}_\mathrm{j}^{-1}.$$

\end{document}